\documentclass[prd,aps,nofootinbib,preprintnumbers]{revtex4}
 \usepackage{amsmath,amssymb,epsf}
 \usepackage{graphicx, pstricks,bbm}

\begin{document}

\title{Non-perturbative decay of udd and QLd flat directions}

\author{A.~Emir~G\"umr\"uk\c{c}\"uo\u{g}lu}
\affiliation{School of Physics and Astronomy, University of
Minnesota, Minneapolis, MN 55455, USA}

\date{\today}

\begin{abstract}
The Minimal Supersymmetric Standard Model has several flat directions, which can naturally be excited during inflation. If they have a slow (perturbative) decay, they may affect the thermalization of the inflaton decay products. In the present paper, we consider the system of udd and QLd flat directions, which breaks the $U(1)\times SU(2)\times SU(3)$  symmetry completely. In the unitary gauge and assuming a general soft breaking mass configuration, we show that for a range of parameters, the background condensate of flat directions can undergo a fast non-perturbative decay, due to non-adiabatic evolution of the eigenstates. We find that both the background evolution and part of the decay can be described accurately by previously studied gauged toy models of flat direction decay.
\end{abstract}

\preprint{arXiv:0910.0854}
\preprint{UMN-TH-2816/09}
\maketitle

\section{Introduction}\label{sec:intro}        
Flat directions are generic features of supersymmetric theories. They are directions in field space along which the renormalizable part of the scalar potential vanishes. The Minimal Supersymmetric Standard Model (MSSM) and its extensions have a plethora of D and F-flat directions \cite{Gherghetta:1995dv},  which are lifted due to supersymmetry breaking. During inflation, if their effective mass is small compared to the Hubble rate, the fields can develop large vacuum expectation values (VEV) along the flat directions of the potential \cite{developvev,Ellis:1987rw}. This growth is bounded above by the non-renormalizable term with lowest dimension, which has the form $\phi^d/M^{d-3}$ with $d \geq 4$. If inflation is long enough, the growth will proceed up to $\langle \phi \rangle \sim (m_\phi M^{d-3})^{1/(d-2)}$ \cite{Dine:1995kz}. If all non-renormalizable terms allowed by gauge invariance are present, each class of flat direction will be lifted by the term with the smallest $d$ allowed by MSSM symmetries \cite{Gherghetta:1995dv}. On the other hand, discrete symmetries may forbid some of such terms and a non-renormalizable term with a higher $d$ may determine the VEV of the flat direction. All possible flat directions in MSSM are excited by terms with $d \le 9$ \footnote{The term at which all flat directions are lifted may be different for the extensions of MSSM. For example, in $\nu$MSSM, no flat direction survives beyond $d=6$ \cite{Basboll:2009tz}.}.  

The formed condensate can have several cosmological implications: In the presence of phase dependent potential terms, the flat directions may source a finite baryon number density through the Affleck-Dine mechanism \cite{Affleck:1984fy, Linde:1985gh, Allahverdi:2000zd}. It has also been suggested that they may be responsible for inflation \cite{Allahverdi:2006iq}. Additionally, due to the large VEV of the flat directions, all the fields coupled to them acquire a large effective mass, slowing down the decays they mediate and resulting in a small perturbative decay rate.  Typically, the perturbative decay of flat direction concludes after $\sim 10^{11}$ rotations \cite{Olive:2006uw}. These long lived flat directions also keep the gauge fields of broken symmetries (assumed to be all the Standard Model ones) heavy, suppressing the scatterings among the inflaton decay products, thus delaying their thermalization \cite{Allahverdi:2005mz}. In addition, the energy density of the (relativistic) inflaton decay products may become sub-dominant over that of (massive) flat directions. The subsequent radiation stage will then be dominated by the thermal distribution of flat direction decay products, rather than those of the inflaton. These effects on thermalization require sufficiently large initial flat direction VEVs, which can be acquired only if non-renormalizable superpotential terms up to $d=11$ are absent \cite{Olive:2006uw}.

However, if the decay of the flat directions is controlled by non-perturbative effects, the effect on thermalization will be very different than the above picture. This possibility was first discussed in \cite{Allahverdi:1999je}, in the framework of a toy model based on F-term type interactions. For this model, the frequencies of the particles coupled to the flat directions evolve adiabatically, not allowing a resonant decay. On the other hand, it was shown in \cite{Olive:2006uw}, that the D-term potential provides non-trivial interactions among the perturbations through a non-diagonal and time dependent mass matrix. Even if the eigenvalues of this matrix evolve adiabatically, the diagonalization procedure itself may be non-adiabatic, due to a fast rotation of the eigenvectors. The resulting exponential decay of the condensate has a much higher rate than the perturbative one, giving a decay after $\mathcal{O}(10)$ rotations of the flat directions. In \cite{Olive:2006uw}, it was also argued that at least two or more flat directions need to be excited for this effect to be realized. The argument is as follows: Since the resonant effect occurs in the D-terms, only the perturbations coupled to the VEVs through the symmetry generators are counted. Out of these degrees of freedom, two per broken symmetry will correspond to a Higgs and a Goldstone. Furthermore, two more (light) degrees of freedom will decouple, corresponding to the real and imaginary parts of fluctuations along each flat direction. In order to have a non-adiabatic mixing, one needs additional light degrees of freedom that the condensate can decay into. To formulate, the number of remaining degrees of freedom present in the system will be
\begin{equation}
\left(
\begin{array}{c}
{\rm remaining}\\
{\rm degrees}
\end{array}
\right)
 = \left(
\begin{array}{c}
{\rm d.o.f.} \\
{\rm in~D~terms}
\end{array}
\right) - 2\,\times \left(
\begin{array}{c}
{\rm broken} \\
{\rm symmetries}
\end{array}
\right)-2\,\times \left(
\begin{array}{c}
{\rm flat} \\
{\rm directions}
\end{array}
\right)\,.
\label{counting}
\end{equation}
As long as this number is zero, there will not be any room for non-perturbative decay. For instance, for the typical cases of single flat directions, no residual degree of freedom is present \cite{Olive:2006uw}. On the other hand, there exist flat directions that are non-exclusive, i.e. they do not give a large mass to each other due to their VEVs. If the conditions to excite a single flat direction are present, one can expect that the whole set of flat directions non-mutually exclusive with that one is excited. If realized, such a case would provide the extra degrees of freedom into which the condensate may decay non-perturbatively.

The longevity of single flat directions was later reiterated by the authors of \cite{Allahverdi:2006xh}, where it was also argued that for the non-perturbative decay of multiple flat directions, one needs some degree of tuning of the initial VEVs: Since different flat directions may be lifted by different non-renormalizable terms in the superpotential \cite{Gherghetta:1995dv}, one may in general expect hierarchical VEVs. Such a case reduces to a single flat direction, which decays only perturbatively. The maximum amount of hierarchy that can provide a non-perturbative decay depends on the ellipticity of the orbits of the VEVs in their complex plane. In later works, gauged toy models with two flat directions \cite{Basboll:2007vt,Gumrukcuoglu:2008fk} and examples from MSSM \cite{Basboll:2008gc} were studied, each verifying that multiple flat directions may decay non-perturbatively. Additionally, in \cite{Gumrukcuoglu:2008fk}, the fast decay was shown to be realized for a range of VEV ratios of three orders of magnitude. This range was found to be a consequence of the phase dependent terms introduced in the fashion of \cite{Affleck:1984fy}.

However, the question of whether the toy models provide a good description of MSSM flat directions needs to be answered. For example, in \cite{Gumrukcuoglu:2008fk}, the gauged toy models of four fields with only $U(1)$ or $SU(2)$ charges have been studied, yet in MSSM, no such flat direction configuration is possible and  generically, for multiple flat directions, the field content has charges of all symmetries. Furthermore, the production of the remaining degrees of freedom (\ref{counting}) may be suppressed if they acquire large masses through the F-terms. Therefore, the main goal of the present work is to find a concrete example from MSSM for which, the decay of the flat directions proceed analogously to the gauged toy model case. The flat directions in the models of \cite{Gumrukcuoglu:2008fk} are decoupled at the background level, and each of the two VEVs evolve independently like two single flat directions. However, having independently evolving VEVs is not a requirement for non-perturbative decay. For instance, flat directions with coupled VEVs also have the necessary ingredients for decay \cite{Basboll:2007vt, Basboll:2008gc}. The latter systems are much more complicated than the former ones and a precise answer requires extended numerical calculation. Our primary focus will be on the system of $u^cu^cd^c$ and $QLd^c$ flat directions and we will show that their decay can be described by the four field toy model of \cite{Gumrukcuoglu:2008fk}.

Additionally, we will address some issues arising from the assumption that the fluctuations along the flat directions are decoupled from the other modes. For instance, for the $QLd^c+LLe^c$ system, Ref. \cite{Basboll:2008gc} claimed that the Higgses and the flat direction perturbations have non-adiabatic mixings. On the other hand, for the toy models of \cite{Gumrukcuoglu:2008fk}, it was shown that the these degrees of freedom indeed decouple from the rest of the action. However, the latter result is a consequence of the assumption that the fields in a given flat direction have equal masses, an assumption not generically applicable to MSSM fields.  If the fields have distinct masses, the flat direction perturbations are no longer decoupled from the Higgses. If these mixings are non-adiabatic, they may result in a non-perturbative decay, even if the counting (\ref{counting}) leaves no extra degrees of freedom. For the models we consider, we show that these mixings have negligible effect and the flat direction perturbations decouple as described in \cite{Olive:2006uw, Gumrukcuoglu:2008fk}. We will first generalize the single flat direction toy model of \cite{Gumrukcuoglu:2008fk} to have arbitrary masses and verify that the flat direction does not decay non-perturbatively. The approximations and methods we adopt in this simple example will provide us the necessary tools for the background evolution of $u^c u^c d^c$ and $Q L d^c$ flat directions, which will also be studied with generic mass terms.

The paper is organized as follows. In Section \ref{sec:mssm}, we discuss different classes of multiple flat directions in MSSM over some examples which allow for non-perturbative decay, and determine which example is most likely to be described by the gauged $4$-field toy model. In Section \ref{sec:formalism}, we review the formalism for the calculations of this decay. In Section \ref{sec:1fd}, we generalize the single flat direction toy model with two complex scalar fields and $U(1)$ gauge field, to include arbitrary soft masses. This provides a basis for non-degenerate mass calculation of a more complicated model, carried out in the next section. In Section \ref{sec:2fd} we present a complete study of $u^cd^cd^c+Q L d^c$ flat directions in MSSM, with arbitrary soft masses, where we compare the final results to the ones for the $4$-field toy model in \cite{Gumrukcuoglu:2008fk}. The results are summarized and discussed in Section \ref{sec:disc}. Finally, we include the technical steps of the calculation in the appendices at the end.


\section{Multiple flat directions in MSSM}\label{sec:mssm}

In this section, we classify some multiple flat direction examples in the MSSM and find a case which has the characteristics of the gauged $4$ field toy model of \cite{Gumrukcuoglu:2008fk}. A classification of multiple flat directions can be made based on the evolution of the VEVs. These are, in the terminology of \cite{Basboll:2008gc}, {\it i) Overlapping flat directions}, where the VEVs are coupled to each other, resulting in a chaotic motion of the phases; {\it ii) Independent flat directions}, where the flat directions are decoupled at the background level and each VEV evolves independently from the others, rotating in an elliptical orbit in their complex plane. For example, the simultaneous excitation of $LLe^c$ and $Q L d^c$ falls into the first category, with the VEV choice 
\begin{equation}
\langle \mu \rangle = \langle \tau^c \rangle = |\Phi|\,{\rm e}^{i\,\sigma} \,,\quad 
\langle d_1 \rangle = \langle s_{\bar{1}}^c \rangle = |\tilde{\Phi}|\,{\rm e}^{i\,\tilde{\sigma}} \,,\quad 
\langle \nu_e \rangle = \sqrt{|\Phi|^2+|\tilde{\Phi}|^2}\,{\rm e}^{i(\sigma+\tilde{\sigma})/2}\,.
\label{LLeQLd}
\end{equation}
On the other hand, as an example for the second class, consider the $LLe^c$ and $u^cd^cd^c$ flat directions, with VEVs,
\begin{equation}
\langle u^c_{\bar{1}}\rangle = \langle s^c_{\bar{2}} \rangle = \langle b^c_{\bar{3}} \rangle= \Phi \,,\quad 
\langle \nu_e\rangle = \langle \mu \rangle = \langle \tau^c \rangle= \tilde{\Phi} \,.
\label{LLeudd}
\end{equation}
Both of these examples were studied in \cite{Basboll:2008gc} where it was shown that only the first case, $LLe^c$ and $Q L d^c$, exhibits the non-adiabatic eigenvector rotation. This result can also be deduced from the counting argument (\ref{counting}): For the $LLe^c$ and $Q L d^c$ example, the field content consists of a squark doublet, a right handed squark, a right handed selectron and two slepton doublets, one of which is shared by the two flat directions, as seen from (\ref{LLeQLd}). The perturbations of these fields contain a total of $32$ real degrees of freedom. On the other hand, the VEV configuration breaks $U(1) \times SU(2) \times SU(3) $ down to $SU(2)$, so the remaining degrees of freedom are $32 - 2 \times 9 - 2\times 2 =10$. On the other hand, for $LLe^c$ and $u^cd^cd^c$, there are $3$ right handed squarks, a right handed selectron and two slepton doublets, so the total real degrees of freedom of the field perturbations is $28$. The VEV configuration (\ref{LLeudd}) breaks $U(1) \times SU(2) \times SU(3)$ completely, so the extra degrees of freedom are $28 - 2 \times 12 - 2 \times 2 = 0$. Hence, for the latter case, there is no room left for the non-adiabatic mixing to occur.

The previously studied multiple flat direction toy models can also be classified based on the above criteria. For instance, the three field model of \cite{Basboll:2007vt}, with D-term potential
\begin{equation}
V = \frac{g^2}{8} \left(\vert \Phi_1\vert^2  -2\, \vert \Phi_2\vert^2 + \vert \Phi_3\vert^2\right)^2\,,
\end{equation}
and VEV configuration
\begin{equation}
\langle \Phi_1 \rangle = \vert \Phi \vert \,{\rm e}^{i\,\sigma} \,,\quad
\langle \Phi_2 \rangle = \frac{\sqrt{\vert \Phi \vert^2 +\vert \tilde{\Phi} \vert^2}}{\sqrt{2}} \,{\rm e}^{i\,(\sigma+\tilde{\sigma})/2} \,,\quad
\langle \Phi_3 \rangle = \vert \tilde{\Phi} \vert \,{\rm e}^{i\,\tilde{\sigma}} \,,\quad
\end{equation}
falls into ``overlapping flat directions'' class, whereas the four field model of \cite{Gumrukcuoglu:2008fk}, with D-term potential
\begin{equation}
V = \frac{g^2}{8} \left(\vert \Phi_1\vert^2  -\vert \Phi_2\vert^2+\vert \Phi_3\vert^2-\vert \Phi_4\vert^2\right)^2\,,
\end{equation}
with VEVs,
\begin{equation}
\langle \Phi_1 \rangle  = \langle \Phi_2 \rangle =\Phi 
\,,\quad
\langle \Phi_3 \rangle  = \langle \Phi_4 \rangle = \tilde{\Phi}\,,
\end{equation}
is a case of ``independent flat directions''. 

We stress here that we expect non-perturbative decay from both systems of ``overlapping'' and ``independent'' VEVs. However, our goal is to find a system of ``independent'' flat directions to be able to use some of the numerical results of the toy model \cite{Gumrukcuoglu:2008fk}. The simplest such example that has extra degrees of freedom is the simultaneous presence of the two flat directions, $Q L d^c$ and $u^cd^cd^c$ which breaks all Standard Model symmetries\footnote{This is welcome, since for the delayed thermalization argument of \cite{Allahverdi:2005mz} to apply, one must assume that all the gauge bosons are heavy.}. The remaining $40- 2\times 12 - 2\times 2 = 12$ degrees of freedom in the spectrum may provide the room needed for non-perturbative decay. Indeed, in the detailed study in Section \ref{sec:2fd}, we verified that these extra degrees participate in a non-adiabatic mixing, resulting in production. 

Of course, the existence of additional light degrees of freedom is not the only requirement for a non-perturbative decay. As summarized in Section \ref{sec:intro}, the ellipticity of the VEV's orbit in its complex plane determines the range of initial VEV ratios for which the rapid production is realized. However, the amount of ellipticity is model dependent. In the case of circular orbits, the non-perturbative decay of the flat directions are highly suppressed if initial VEVs are not comparable. On the other hand, for a model with vanishing superpotential, the motion is purely radial \cite{Dine:1995kz,Giudice:2008gu} and even a single flat direction may undergo a fast decay. For these considerations, in the toy models of \cite{Gumrukcuoglu:2008fk}, the choice of \cite{Affleck:1984fy} which results in an intermediate ellipticity was adopted, where quartic terms of the form $\lambda (\Phi_1^2 \Phi_2^2 +{\rm c.c})$ with $\lambda \propto m^2/|\Phi_0|^2$ provide the angular momentum. Due to the complexity of the MSSM example, we will assume that an ellipticity of the same order was acquired initially and disregard such terms afterwards, since they quickly become sub-dominant over the mass terms once the fields start moving, and their amplitudes decrease (due to the expansion of the universe).

\section{Formalism}\label{sec:formalism}
In this section, we summarize the formalism of \cite{Nilles:2001fg} and derive the conditions under which a production is expected.
We start from the action of $N$ coupled scalars in Minkowski space in the form
\begin{equation}
\mathcal{S} = \frac{1}{2}\int d^3k \,	d\eta\,\left[ \Psi^{\prime \, \dagger}  \Psi' - \Psi^\dagger \,\Omega^2(\eta)\Psi\right]\,,
\label{canform}
\end{equation}
where $\Psi$ is an $N$ dimensional vector and $\Omega^2$ is a time dependent $N\times N$ real matrix. We define $C$ to be the orthogonal matrix that diagonalizes $\Omega^2$ through
\begin{equation}
C^T \Omega^2  C  = \omega^2\;({\rm diagonal})\,.
\end{equation}
It is crucial to note that in general, the matrix $C$ is time dependent. As a consequence, the fields in the new basis $\tilde{\Psi} \equiv C^T \Psi$ are {\it not} the physical eigenstates of the system as long as the matrix $C$ evolves non-adiabatically. The quantum evolution equations for the generalized Bogolyubov coefficients are  \cite{Nilles:2001fg},
\begin{eqnarray}
\alpha' &=& \left(-i\,\omega-I\right)\,\alpha + \left(\frac{\omega'}{2\,\omega} - J\right)\,\beta\,,\nonumber\\
\beta' &=& \left(i\,\omega-I\right)\,\alpha + \left(\frac{\omega'}{2\,\omega} - J\right)\,\alpha\,,
\label{bogoleq}
\end{eqnarray}
with
\begin{equation}
I,\,J = \frac{1}{2} \left(\sqrt{\omega}\,\Gamma \,\frac{1}{\sqrt{\omega}}\pm \frac{1}{\sqrt{\omega}}\,\Gamma \,\sqrt{\omega}\right) \,, \quad\quad
\Gamma = C^T C'\,,
\label{eqij}
\end{equation}
where, by construction, $\Gamma$ and $I$ are anti-symmetric, whereas $J$ is symmetric. Additionally, the canonical commutation relations impose the conditions
\begin{equation}
\alpha\,\alpha^\dagger - \beta^\star\,\beta^T = \mathbbm{1}\,,\quad\quad
\alpha\,\beta^\dagger - \beta^\star\,\alpha^T = 0\,.
\end{equation}
Finally, the occupation number of $i$--th physical state is
\begin{equation}
n_i  = \left(\beta^\star\,\beta^T\right)_{ii}\;({\rm no~sum})\,.
\end{equation}

The evolution equations can be seen to be constructed out of two contributions. The first part consists of the anti-Hermitian matrices $(\pm i\,\omega -I)$. As such a matrix has purely imaginary eigenvalues, the effect of this part is to rotate produced particles of physical states into each other, while preserving the total occupation number ${\rm Tr}(\beta^\star\beta^T)$. The second part however is the Hermitian matrix $(\omega'/2\omega-J)$ which causes the change in the occupation numbers, thus responsible for any potential production. Following \cite{Gumrukcuoglu:2008fk}, we define the ``adiabaticity matrix'' $\mathcal{A}$ as
\begin{equation}
\mathcal{A} \equiv \frac{\omega'}{\omega^2} - \frac{1}{\sqrt{\omega}}(2\,J)\frac{1}{\sqrt{\omega}} = \frac{\omega'}{\omega^2} - \left(\Gamma \,\frac{1}{\omega} - \frac{1}{\omega}\,\Gamma\right)\,.
\label{adia}
\end{equation}
In order to have particle production through parametric resonance, at least one component of the matrix $\mathcal{A}$ should satisfy \cite{Gumrukcuoglu:2008fk}
\begin{equation}
|\mathcal{A}_{ij}| \gtrsim 1 \,.
\label{adiabatic}
\end{equation}
The diagonal elements of this condition
\begin{equation}
\left\vert\frac{\omega_i'}{\omega_i^2}\right\vert >1\;\;({\rm no~sum})\,,
\label{adiadiag}
\end{equation}
requires that the eigenfrequencies evolve non-adiabatically. Notice that this is the standard condition for non-adiabatic evolution, valid for the case in which the produced fields are not mixed.  On the other hand, the off diagonal components
\begin{equation}
\left \vert\Gamma_{ij} \left(\frac{1}{\omega_i}-\frac{1}{\omega_j}\right)\right\vert >1 \;\;({\rm no~sum})\,,
\label{adiaoffd}
\end{equation}
measure the non-adiabatic evolution of the eigenvectors corresponding to the physical modes. It should be noted that the off-diagonal components corresponding to two degenerate states correctly vanish, as the rate of change of the rotation between their corresponding eigenvectors is not physical\footnote{At the quadratic level, with all the interaction disregarded, fields of equal mass are identical.}. We emphasize that the condition (\ref{adiabatic}) is {\it not} a sufficient condition for production; a detailed numerical analysis is needed to correctly determine if non-perturbative decay occurs.

Throughout the paper, we exploit the large hierarchy between the TeV scale soft masses and the VEVs, by performing a series expansion in the ratio of these scales, which we denote by $\epsilon$. The physical modes are expected to be either heavy or light, with eigenfrequencies of order $\mathcal{O}(\epsilon^0)$ or $\mathcal{O}(\epsilon^1)$, respectively. The scale of the rate of change for the background quantities are typically of order of soft masses, so differentiation of these with respect to time raises the order of $\epsilon$ by one. As the matrix $C$ which diagonalizes the frequency matrix is unitary, its leading order in the expansion is the $\epsilon^0$ term. This implies that
\begin{equation}
\Gamma = \mathcal{O}(\epsilon)\,.
\end{equation}
From (\ref{adiaoffd}), we see that $\mathcal{A}_{ij}$ is of order $\epsilon$ or higher, for
\renewcommand{\labelenumi}{\it \roman{enumi}.}
\begin{enumerate}
\item rotations between two heavy modes;
\item rotations between states that are degenerate at least at the leading order;
\item changes in the frequency of a heavy mode.
\end{enumerate}
That is, for the above cases, condition (\ref{adiabatic}) cannot be satisfied. The only components of the adiabaticity matrix that can be of order $\epsilon^0$ are the diagonal elements corresponding to the light modes, and off diagonal elements corresponding to rotations between a light mode and another mode which may either be light or heavy. In the latter case, the frequencies of the two modes need to be different at leading order. Terms of order $\epsilon^2$ in any component of $\Gamma$ matrix lead to negligible $\mathcal{A}$ contribution which does not change the picture and will not be calculated in this work.


\section{2-field toy model revisited: Non-degenerate mass case}\label{sec:1fd}

In this section, we consider the toy model with two complex scalar fields of opposite $U(1)$ charges and study the effect of arbitrary soft breaking masses. Since we expect a new mixing between the light flat direction fluctuations and the Higgs, we wish to verify that this new coupling does not contribute to non-perturbative decay. The approximations and methods adopted here will also form a basis for the background calculations of a more complicated MSSM example, studied in Section \ref{sec:2fd}.

We start by generalizing the single flat direction potential of \cite{Gumrukcuoglu:2008fk} to arbitrary masses,
\begin{equation}
V= m_1^2 |\phi_1|^2 + m_2^2 |\phi_2|^2 + \lambda (\phi_1^2 \phi_2^2 +{\rm c.c.}) + \frac{e^2}{8} \left(q_1 |\phi_1|^2 +q_2 |\phi_2|^2\right)^2\,,
\label{pot2fld}
\end{equation}
where $e$ is the coupling constant of $U(1)$ gauge and the $U(1)$ charges of the fields are $q_1 = -q_2 = 1$. The quartic $\lambda$ term is assumed to be $\propto \frac{m_1^2+m_2^2}{|\Phi_0|^2}$, a choice compatible with the Affleck-Dine scenario. Here, $\Phi_0$ is the value of the flat direction VEV when it is still frozen. The above potential reduces to the one considered in \cite{Gumrukcuoglu:2008fk} in the limit $m_1=m_2$.

The complete action is
\begin{equation}
S = \int d^4 x \,\sqrt{-g} \left[D_\mu \phi_1 D^\mu \phi_1^\dagger + D_\mu \phi_2 D^\mu \phi_2^\dagger - \frac{1}{4} F_{\mu \nu} F^{\mu \nu} - V\right]\,,
\end{equation}
with the covariant derivatives
\begin{equation}
D_\mu \phi_i \equiv \left(\partial_\mu - \frac{i\,e}{2}\,q_i A_\mu\right)\phi_i\,.
\end{equation}
Here and in the remainder of the text, we use the metric in conformal time $d \eta \equiv R(t) dt$
\begin{equation}
g_{\mu \nu} = R^2\,{\rm diag} \left( 1 , \;-1,\;-1,\;-1\right)\,,
\end{equation}
where $R$ is the scale factor.

\subsection{Background}\label{sec:1fdBG}
We decompose the background fields as
\begin{equation}
\langle \phi_1 \rangle= \frac{F+\Delta F}{2\,R}\,{\rm e}^{i\,(\Sigma + \Delta \Sigma)}
\quad\,,\quad
\langle \phi_2 \rangle= \frac{F-\Delta F}{2\,R}\,{\rm e}^{i\,(\Sigma - \Delta \Sigma)}\,,
\end{equation}
where we parametrized the two phases with their sum ($2\,\Sigma$) and their difference ($2\,\Delta \Sigma$). Since the two fields have opposite charge, the $U(1)$ transformation affects only the phase difference, which we fix at $\Delta \Sigma = 0$. In the degenerate mass case, $m_1=m_2$, the D-flatness condition $|\phi_1|=|\phi_2|$ allows us to have $\Delta F =0$. However, as we show below, in the general case, exact D-flatness cannot be attained and we need to introduce a non-zero $\Delta F$.

For convenience, we assume that the initial gauge field vanishes ($\langle A_i\rangle=0$) (which is kept at zero by the equations of motion also at later times). With these considerations, the Maxwell's equations for $\langle A_\mu \rangle$ can be reduced to the constraint
\begin{equation}
\langle A_0 \rangle= \frac{4}{e}\,\frac{F\,\Delta F\,\Sigma'}{F^2+\Delta F^2}\,,
\label{constA0}
\end{equation}
whereas the background equations of motion are
\begin{eqnarray}
&&F'' + \left(m^2\,R^2-\frac{R''}{R}-\Sigma'^2\right) F +\frac{\lambda}{2}\,F^3\,\cos(4\,\Sigma) + \delta m^2 R^2 \Delta F
\nonumber\\
&&\quad\quad\quad
\quad\quad\quad
\quad\quad\quad\quad\quad\quad
+ \frac{1}{4} F\,\Delta F^2\,\left(e^2-2\,\lambda\,\cos(4\,\Sigma) + \frac{16\,\Delta F^2\,\Sigma'^2}{(F^2+\Delta F^2)^2}\right) =0\,,
\nonumber\\
&&\Delta F''+\left(m^2\,R^2-\frac{R''}{R}-\Sigma'^2\right) \Delta F+\frac{\lambda}{2}\,\Delta F^3\,\cos(4\,\Sigma) + \delta m^2 R^2 F
\nonumber\\
&&\quad\quad\quad
\quad\quad\quad
\quad\quad\quad\quad\quad\quad
+ \frac{1}{4} F^2\,\Delta F\,\left(e^2-2\,\lambda\,\cos(4\,\Sigma) + \frac{16\,F^2\,\Sigma'^2}{(F^2+\Delta F^2)^2}\right) =0\,,
\nonumber\\
&&\left(\frac{(F^2-\Delta F^2)^2}{F^2+\Delta F^2}\Sigma'\right)' - \frac{\lambda}{2}\,\left(F^2-\Delta F^2\right)^2\,\sin(4\,\Sigma) =0\,.
\label{eom1f-bg}
\end{eqnarray}
where we defined $m^2 \equiv (m_1^2 + m_2^2)/2$ , $\delta m^2 \equiv (m_1^2 - m_2^2)/2$ and $\delta m^2$ can also be negative. Throughout the paper, a prime denotes differentiation with respect to the conformal time. From the second of (\ref{eom1f-bg}), we see that the D-flat solution $\Delta F=0$ is allowed only if $\delta m^2 =0$. 

Although the above equations are all we need to solve the background evolution, it is very useful to write them in a series approximation, to quantify the 
modifications due to the introduction of a non-zero mass difference $\delta m^2$.  We will work in the limit
\begin{equation}
\left\{ m\,R \,, \Sigma' \,, \frac{F'}{F} \right\} \ll F\,.
\label{1fdapp}
\end{equation}
In other words, we expand the equations of motion (\ref{eom1f-bg}) in terms of TeV$/$VEV ratio. For bookkeeping, we denote the order of expansion by $\epsilon$. In this fashion, we expand the background quantities as a power series in $\epsilon$
\begin{equation}
Q= \sum_i\, Q_i \epsilon^i\,,
\end{equation}
where $Q$ can be $F$, $\Sigma$ or $\Delta F$.  Recalling that we chose $\lambda \propto m^2/F^2$, its order will be $\epsilon^2$. 

\begin{figure}[ht]
\centerline{
\includegraphics[width=0.6\textwidth,angle=-90]{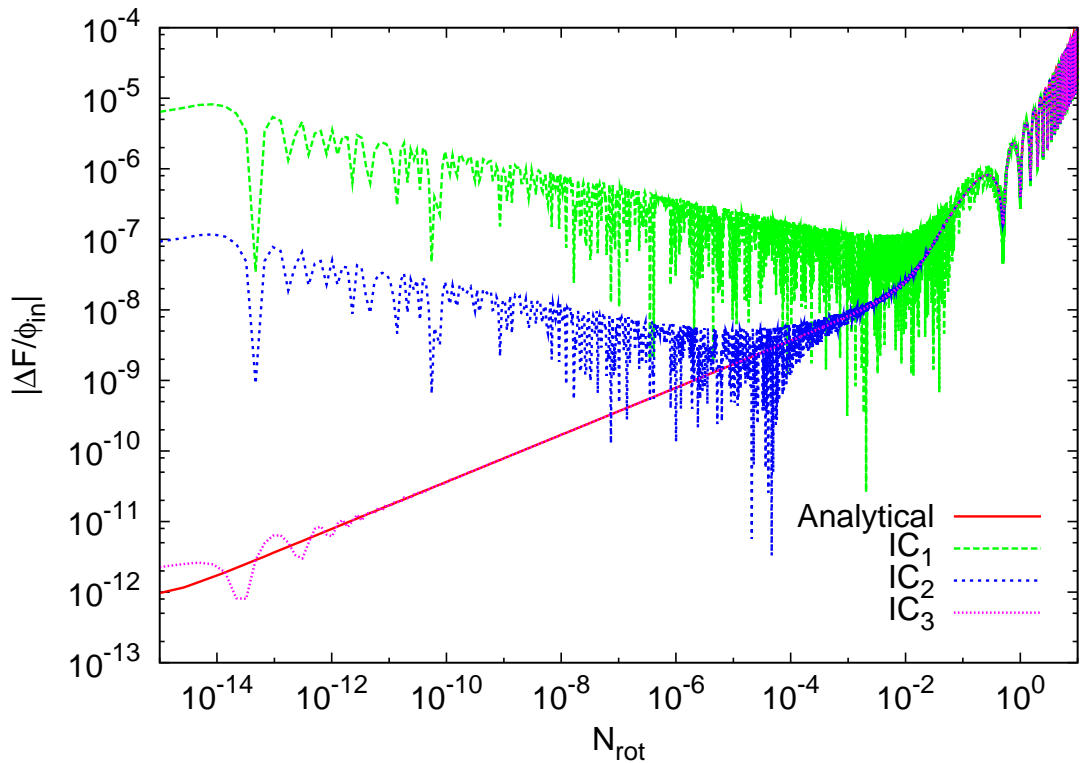}
}
\caption{
The evolution of $\Delta F$ for the first $10$ rotations of the flat direction. For this plot, we evolved the exact equations (\ref{eom1f-bg}), and in all cases shown, the parameters are $m = 10^{-6}\, e\,|\phi_{\rm in}|$, $\delta m = m/2$. The curves correspond to three different initial conditions for $\Delta F$, where $IC_1$ is $\Delta F_{\rm in} = 7\, \delta m/e$, $IC_2$ is $\Delta F_{\rm in} = 10^{-1} \delta m/e$ and $IC_3$ is $\Delta F_{\rm in} = 5\,\delta m^2/e^2 /|\phi_{\rm in}|$. We also present the plot of the analytic result of (\ref{hexp}) for comparison, which is consistent with initial conditions $\Delta F_{\rm in} = -2\,\delta m^2/e^2 /|\phi_{\rm in}|$. For the cases $IC_1$ and $IC_2$, the leading order term in $\Delta F_{\rm in}$ is $\mathcal{O}(\epsilon)$, which becomes sub-dominant over the second order term within a tenth of a rotation.
}
\label{approxi} 
\end{figure}

Determining the leading order terms of $\epsilon$ expansion in the time derivatives is less trivial. The term $\frac{e^2}{4} F\,\Delta F^2$ in the equation of motion for $F$ and the term $\frac{e^2}{4} F^2\,\Delta F$ in the equation for $\Delta F$ may result in oscillations with frequencies of order VEV. On the other hand, the rotation of the flat directions, which has a TeV scale frequency, is only sensitive to the average of the fast oscillations. Therefore, we carry out this averaging formally, that is, we assume that all the fields evolve with the small mass scale, so each time derivative increases the order of the quantity by $\epsilon$. This way, we write the equations of motion at each order of expansion. The equations (\ref{eom1f-bg}) at order $\mathcal{O}(\epsilon^0)$ give two algebraic relations
\begin{equation}
\frac{e^2}{4}\,F_0\,\Delta F_0^2 = 0 \,,\quad\quad
\frac{e^2}{4}\,F_0^2\,\Delta F_0 = 0 \,,
\end{equation}
which are solved by $\Delta F_0=0$. When this solution is plugged in, the only $\mathcal{O}(\epsilon^1)$ equation is
\begin{equation}
\frac{e^2}{4}\,F_0^2\,\Delta F_1 = 0 \,,
\end{equation}
which removes the $\mathcal{O}(\epsilon^1)$ term of $\Delta F$. Next, we write down $\mathcal{O}(\epsilon^2)$ equations,
\begin{eqnarray}
&& F_0'' + F_0 \left[m^2\,R^2 -\Sigma_0'^2-\frac{R''}{R}+ \frac{\lambda}{2}\,F_0^2 \cos(4\,\Sigma_0)\right] = 0\,,\nonumber\\
&&\frac{F_0}{4}\,\left(e^2\,F_0\,\Delta F_2 + 4\,\delta m^2 R^2\right) = 0 \,,\nonumber\\
&&\left(F_0^2\,\Sigma_0'\right)' - \frac{\lambda}{2}\,F_0^4\sin(4\,\Sigma_0) = 0\,,
\end{eqnarray}
where the equation for $\Delta F_2$ is again an algebraic one. Finally, the $\mathcal{O}(\epsilon^3)$ equations are
\begin{eqnarray}
&&F_1''+ F_1 \left(m^2 R^2 -\Sigma_0'^2 -\frac{R''}{R}\right) -2\,F_0 \Sigma_0' \Sigma_1' + \frac{\lambda}{2} F_0^2 \left[ 3\,F_1 \cos(4\,\Sigma_0) - 4\,F_0\Sigma_1 \sin(4\,\Sigma_0)\right]=0\,,\nonumber\\
&&\frac{e^2}{4} F_0^2 \Delta F_3 - \delta m^2 R^2 F_1 = 0 \,,\nonumber\\
&&\left(F_0^2\Sigma'_1\right)' + 2\,\left(F_0 F_1'-F_1 F_0'\right)\Sigma_0'-\lambda F_0^3\left[F_1 \sin(4\,\Sigma_0)+2\,F_0 \Sigma_1 \cos(4\,\Sigma_0)\right]=0\,.
\end{eqnarray}
Collecting these solutions, we find
\begin{eqnarray}
\epsilon^2 \left[F_0''+\epsilon F_1'' +\mathcal{O}(\epsilon^2)\right] &=& - \left[\left(m^2\,R^2 - \frac{R''}{R} - \Sigma_0'^2\right)F_0  + \frac{\lambda}{2}\,F_0^3\,\cos(4\,\Sigma_0)\right] \epsilon^2
\nonumber\\
&&\times\left[1+\left(\frac{F_1}{F_0}+
\frac{\lambda\,F_0\,F_1 \cos(4\,\Sigma_0)-2\,\lambda\,F_0^2\,\Sigma_1\,\sin(4\,\Sigma_0) -2\,\Sigma_0'\,\Sigma_1'}{\frac{\lambda}{2}\,F_0^2\cos(4\,\Sigma_0) +m^2\,R^2-\Sigma_0'^2- \frac{R''}{R}}\right)\epsilon+ \mathcal{O}(\epsilon^2)\right]\,,
\nonumber\\
\epsilon^2 \left[\Sigma_0''+\epsilon \Sigma_1'' +\mathcal{O}(\epsilon^2)\right] &=& \left[\frac{\lambda}{2}\,F_0^2\,\sin(4\,\Sigma_0)-\frac{2\,F_0'\,\Sigma_0'}{F_0}\right]\epsilon^2
\nonumber\\
&&\times\left[
1+
2\,\left(\frac{F_1}{F_0}+
\frac{
 \lambda\,F_0^2\,\Sigma_1\,\cos(4\,\Sigma_0) + \frac{\Sigma_0'}{F_0^2}\,\left(3\,F_0'\,F_1 - F_0\,F_1'\right) -\frac{F_0'}{F_0}\,\Sigma_1'
}{\frac{\lambda}{2}\,F_0^2\,\sin(4\,\Sigma_0)-2\,\frac{F_0'}{F_0}\,\Sigma_0'}\right)\epsilon+ \mathcal{O}(\epsilon^2)
\right]\,,
\label{ddexp}
\end{eqnarray}
where the leading order equations match with those of the degenerate mass case \cite{Gumrukcuoglu:2008fk}. On the other hand, the solution for $\Delta F$ is
\begin{eqnarray}
\Delta F  = -\frac{4\,\delta m^2\,R^2}{e^2\,F_0}\,\epsilon^2 \left(1 -\frac{F_1}{F_0} \epsilon+ \mathcal{O}(\epsilon)^2\right)\,.
\label{hexp}
\end{eqnarray}
This solution guarantees that at every order computed, the large mass contribution for $\Delta F$ cancels with the term $\delta m^2 R^2 F$. To justify the assumption of neglecting the fast oscillating part of $\Delta F$, we show in Figure \ref{approxi} the numerical evolution of $\Delta F$, compared with the approximate result (\ref{hexp}). We evolve the exact equations of motion (\ref{eom1f-bg}) numerically, giving several initial conditions for $\Delta F$ that are inconsistent with (\ref{hexp}). We find that even for initial conditions implying $\Delta F_1 \neq 0$, the evolution of $\Delta F$ converges to (\ref{hexp}) within a tenth of a rotation, after which the approximate solution is valid. On the other hand, we found that the leading order terms in equations (\ref{ddexp}) describe the evolution correctly as long as $\Delta F_0 = 0$. This latter assumption can be justified by looking at the D-term, which reads
\begin{equation}
V_D = \frac{e^2}{8\,R^4}\,F^2\,\Delta F^2 \,.
\end{equation}
Unless the leading order term in $\Delta F$ is at least of $\mathcal{O}(\epsilon)$, the D-term will be the dominant contribution in the potential. Conversely, for the case where the D-term is at least comparable to the mass terms in the potential, the flat direction is more flat than the previous case and is a more preferred configuration. For the solution (\ref{hexp}), the D-term potential is, 
\begin{equation}
V_D = \frac{2\,\delta m^4}{e^2} \epsilon^4 + \mathcal{O}(\epsilon^5)\,,
\end{equation}
which gives a negligible contribution to the potential of the flat direction.

Finally, we note that the two terms in the magnitude of the VEV, i.e. $F_0$ and $\Delta F_2$, redshift with different powers of the scale factor. Specifically, once the field starts to move at $H \sim m$ (or $R = R_\phi$), $F_0$ will evolve as
\begin{equation}
F_0 \sim 2\,|\phi_{\rm in}| \frac{R_\phi^{3/2}}{\sqrt{R}} \,,\;({\rm  for~} H<m)\,,
\end{equation}
where we denote the initial value of the VEV as $|\phi_{\rm in}|$. Therefore, it is important to check the range of validity of our approach. The ratio of the two constituents is
\begin{equation}
\left\vert\frac{\Delta F_2}{F_0 } \right\vert \sim \frac{\delta m^2}{e^2 \vert\phi_{\rm in}\vert^2} \left(\frac{R}{R_\phi}\right)^3\,.
\label{FoverdF}
\end{equation}
Assuming $\delta m\sim {\rm TeV}$, $\vert\phi_{\rm in}\vert \sim 10^{-2} M_p$ and $e^2 \sim 0.1$, this ratio will become of order one after about $20$ e-folds of expansion from the time when the field starts to move. Since this corresponds to $10^{12}$ rotations of the flat direction in a matter dominated background (or $10^{16}$ rotations if radiation dominates), we can safely use the expressions (\ref{ddexp}) and (\ref{hexp}) in the rest of the calculation. The time at which  (\ref{FoverdF}) becomes of order one also corresponds to the time when the D-term becomes comparable to the mass terms, so in the antecedent evolution, the potential of the flat direction will be dominated by the mass terms, just like in the degenerate mass case.

\subsection{Perturbations}\label{sec:1fdPER}
After specifying to unitary gauge by fixing the gauge variant degrees to zero, the perturbations can be decomposed in terms of gauge invariant degrees of freedom as
\begin{equation}
\delta \phi_1 = \frac{r+\delta_H + i\,(F+\Delta F)\sigma}{2\,R}\,{\rm e}^{i\,\Sigma}
\quad\,,\quad
\delta \phi_2 = \frac{r-\delta_H + i\,(F-\Delta F)\sigma}{2\,R}\,{\rm e}^{i\,\Sigma}
\end{equation}
where $r$, $\delta H$ and $\sigma$ are perturbations of $F$, $\Delta F$ and $\Sigma$, respectively. In the limit of degenerate masses, it was shown in \cite{Gumrukcuoglu:2008fk} that the two real degrees of freedom in $\delta \phi_1 +\delta\phi_2$, which correspond to the fluctuations along the flat direction were decoupled from the rest and did not give any contribution to non-perturbative decay. Additionally, $\delta \phi_1 - \delta\phi_2$ was identified as the physical Higgs, which only mixed with the (also heavy) longitudinal component of the $U(1)$ gauge field, and did not result in any non-adiabatic effect. On the other hand, we will see that the presence of a non-zero $\Delta F$ will cause the light modes corresponding to the flat direction perturbations to mix with the other degrees of freedom. Here, we will summarize the results of the detailed calculation of the action, presented in Appendix \ref{AppB}. 

It is convenient to decompose the vector field into transverse and longitudinal parts through
\begin{equation}
A_i = A_i^T+\partial_i L\,, \quad{\rm with~~~}\partial_i A_i^T = 0.
\end{equation}
The two transverse vector components decouple at the linear level, with mass
\begin{equation}
m^2_{\rm gauge} = \frac{e^2\,\left(F^2+\Delta F^2\right)}{4 \, R^2}
\label{mvecU1}
\end{equation}
Unlike the case with degenerate masses, the flat direction perturbations $r$ and $\sigma$ do not decouple from the rest. As a consequence, we end up with a coupled system of $r$, $\sigma$, $\delta_H$, $L$, along with the non-dynamical perturbations of $A_0$. After integrating it out and redefining the fields to have canonical kinetic terms, we end up with an action of the form (\ref{canform}). Disregarding the quickly suppressed terms proportional to $R''$ and $\lambda$, we find that the eigenmasses of the physical modes are, in the approximation (\ref{1fdapp}),
\begin{eqnarray}
m_A^2 &=& \left[m^2\right] +\mathcal{O}(\epsilon^3)
\,,\nonumber\\
m_B^2 &=& \left[m^2\right]+\mathcal{O}(\epsilon^3)
\,,\nonumber\\
m_C^2 &=& \frac{e^2 F_0^2}{4\,R^2}\, + \left\{ \frac{e^2}{2\,R^2}\,F_0\,F_1\right\} +\left[\frac{e^2}{4\,R^2}\left(F_1^2+2\,F_0\,F_2\right)\right]+\mathcal{O}(\epsilon^3)
\,,\nonumber\\
m_D^2 &=& \frac{e^2 F_0^2}{4\,R^2} + \left\{ \frac{e^2}{2\,R^2}\,F_0\,F_1\right\} +\left[\frac{e^2}{4\,R^2}\left(F_1^2+2\,F_0\,F_2\right)+m^2+ \frac{3\ \Sigma_0'^2}{R^2}\right]+\mathcal{O}(\epsilon^3)\,,
\label{eigenmass1fd}
\end{eqnarray}
where the terms outside parenthesis, in curly parenthesis and in square brackets are, respectively, of zeroth, first and second order in $\epsilon$. We identify the heavy mode $(C)$ as the longitudinal vector component, as its mass coincides with that of the transverse components (\ref{mvecU1}) at the order shown. The other heavy mode $(D)$ then corresponds to the physical Higgs. The light modes $(A,B)$ can be identified as the excitations along the flat direction. All four eigenmasses vary adiabatically and the rotation between the modes is suppressed by high powers of $\epsilon$.

This concludes the study of the single flat direction toy model with two fields, generalized to have arbitrary masses. We have shown that the mixing between the (light) flat direction perturbations and the Higgs does not provide a quick rotation, thus verifying the results discussed in \cite{Olive:2006uw} and computed in \cite{Gumrukcuoglu:2008fk}. Once the TeV/VEV expansion is applied, the equations for the degenerate mass case \cite{Gumrukcuoglu:2008fk} are recovered up to sub-dominant (and negligible) terms.
%


\section{$u^cd^cd^c+QLd^c$ flat directions}\label{sec:2fd}

In this section, we consider the simultaneous excitation of $u^cd^cd^c$ and $QLd^c$ flat directions, which breaks $U(1)\times SU(2)\times SU(3)$ completely. There are much simpler examples in MSSM with less degrees of freedom and/or broken symmetries, but as discussed in Section \ref{sec:mssm}, this is the simplest case of two ``independent flat directions'' which has sufficient degrees of freedom for non-perturbative decay.
\begin{table*}
\begin{tabular}{|c||c|c|c|}
\hline
&Field
&  Y (hypercharge)&
degree of freedom  \\
\hline \hline
$\phi_1$ &  $u^c$ & $-\frac{4}{3}$ & 6\\\hline
$\phi_2$ &  $s^c$ & $\frac{2}{3}$ & 6\\\hline
$\phi_3$ &  $b^c$ & $\frac{2}{3}$ & 6\\\hline\hline
$\phi_4$ &  $d^c$ & $\frac{2}{3}$ & 6\\\hline
$\phi_5$ &  $L_{\rm e}$ & $-1$ & 4\\\hline
$\phi_6$ &  $Q_{\rm c/s}$ & $\frac{1}{3}$ & 12\\\hline\hline
\end{tabular}%
\caption{Summary table of notation and hypercharges of the field content. }
\label{tab1}
\end{table*}

As it is apparent from the naming scheme of the flat directions, the fields that acquire VEVs in this example are a squark doublet, four right handed squarks and a slepton doublet. In Table \ref{tab1}, we show the definitions for the fields $\phi_i$ as well as the choices of the specific flavors for each particle. The D-term potential for this field content is,
\begin{equation}
\frac{1}{2} D^2 = \frac{1}{8}\, g_1^2 \left\vert \sum_{i=1}^6 \phi_i^\dagger Y \phi_i \right\vert^2
+\frac{1}{8}\, g_2^2 \sum_{a=1}^3 \left\vert \sum_{i=1}^6 \phi_i^\dagger \sigma_a \phi_i \right\vert^2
+\frac{1}{8}\, g_3^2 \sum_{a=1}^8 \left\vert \sum_{i=1}^6 \phi_i^\dagger \lambda_a \phi_i \right\vert^2\,,
\end{equation}
where $g_1$, $g_2$, $g_3$ are the coupling constants, $Y$, $\sigma_a$, $\lambda_a$ are the generators of $U(1)$, $SU(2)$ and $SU(3)$ groups respectively, with $SU(2)$ and $SU(3)$ generators represented by Pauli and Gell-Mann matrices (see e.g. \cite{Itzykson:1980rh}). On the other hand, the only F-term potential that gives non-zero contribution for the choices of flavors in Table \ref{tab1} is
\begin{equation}
\vert F \vert^2 = |y_d \, \phi_6\,\phi_2|^2 \,,
\end{equation}
where $y_d$ is the Yukawa coupling in the superpotential term $y_d \,Q\cdot H_d \,d^c$. 

A VEV configuration which is both F and D-flat can be chosen as
\begin{equation}
\langle u_{\bar{1}}^c \rangle = \langle s_{\bar{2}}^c \rangle = \langle b_{\bar{3}}^c \rangle = \Phi \,,\quad
\langle d_{\bar{1}}^c \rangle =\langle \nu_e \rangle =\langle s_1 \rangle = \tilde{\Phi}  \,.
\label{vevchoice-degen}
\end{equation}
where the subscripts $1,2,3$ denote the $SU(3)$ color charges. The $12$ degrees of freedom for the $6$ components have been reduced to $4$ by using the $4$ D-flatness conditions and $4$ gauge freedoms corresponding to diagonal generators. Since the $SU(3)$ charges of the components of $s^c$ and $Q$ which have VEVs are different, the choice is F-flat. Furthermore, the total $U(1)$, $SU(2)$ and $SU(3)$ charges of the components with same VEVs cancel, so the D-term also vanishes. 

Now, we introduce soft supersymmetry breaking mass terms in the potential,
\begin{equation}
V_{\rm soft} = \sum_i m^2_{i} \phi_i^\dagger \phi_i\,.
\end{equation}
If the mass configuration has a degeneracy such that $m_1=m_2=m_3$ and $m_4=m_5=m_6$, the equations of motion for the background is then identical to the one in \cite{Gumrukcuoglu:2008fk} for the four field toy model, with the exception of the phase dependent terms that provide the rotation. However, as we discussed earlier, 
such a mass setup implies that different ingredients, such as the sleptons and squarks have equal masses. In the rest of the calculation, we will assume a general mass configuration in the fashion of Section \ref{sec:1fd}. Generalizing the configuration (\ref{vevchoice-degen}), we have
\begin{equation}
\langle u_{\bar{1}}^c \rangle = \Phi_1 \,,\quad
\langle s_{\bar{2}}^c \rangle = \Phi_2 \,,\quad
\langle b_{\bar{3}}^c \rangle = \Phi_3 \,,\quad
\langle d_{\bar{1}}^c \rangle = \Phi_4 \,,\quad
\langle \nu_e \rangle = \Phi_5 \,,\quad
\langle s_1 \rangle = \Phi_6 \,,
\label{vevchoice}
\end{equation}
The total potential that we will use is,
\begin{equation}
V = |F|^2 + \frac{1}{2} D^2 + V_{\rm soft}\,,
\label{potpot}
\end{equation}
where we omit the phase dependent quartic soft terms. However, we will assume that an initial angular momentum was already provided after the flat directions started to evolve and their source became quickly sub-dominant afterwards. The configuration (\ref{vevchoice}) is still F-flat. On the other hand, as we will show, the D-flatness cannot be obtained exactly and we can only remove $4$ degrees of freedom in the background by applying gauge transformations with diagonal generators. Therefore, the VEV configuration has $8$ degrees of freedom. In the following subsections, we study the background and perturbations separately and the technical details for these calculations are summarized in Appendices \ref{AppC} and \ref{AppD}.

\subsection{The model and VEV configuration}\label{sec:2fdBG}

We introduce a general decomposition of the background fields in the unitary gauge as
\begin{eqnarray}
\Phi_1 &=& \frac{F+2\,\Delta F_1}{\sqrt{6}\,R}\,{\rm e}^{i\,\Sigma}\,,\quad
\Phi_2 = \frac{F - \Delta F_1 +\Delta F_2}{\sqrt{6}\,R}\,{\rm e}^{i\,\Sigma} \,,\quad
\Phi_3 = \frac{F - \Delta F_1 -\Delta F_2}{\sqrt{6}\,R}\,{\rm e}^{i\,\Sigma} \,,\nonumber\\
\Phi_4 &=& \frac{G+2\,\Delta G_1}{\sqrt{6}\,R}\,{\rm e}^{i\,\tilde{\Sigma}} \,,\quad
\Phi_5 = \frac{G - \Delta G_1 +\Delta G_2}{\sqrt{6}\,R}\,{\rm e}^{i\,\tilde{\Sigma}} \,,\quad
\Phi_6 = \frac{G - \Delta G_1 -\Delta G_2}{\sqrt{6}\,R}\,{\rm e}^{i\,\tilde{\Sigma}} \,.
\end{eqnarray}
Next, we integrate out the non-dynamical temporal components of all the gauge fields. The resulting equations of motion are rather bulky and are discussed in detail in Appendix \ref{AppC}. Here, we present them in the TeV/VEV expansion. First, we apply the two flat direction analogue of the approximations (\ref{1fdapp}),
\begin{equation}
\left\{ m_i\,R \,, \Sigma' \,, \tilde{\Sigma}'\,, \frac{F'}{\sqrt{F^2+G^2}}\,, \frac{G'}{\sqrt{F^2+G^2}} \right\} \ll \left\{F\,, G \right\}\,.
\label{2fdapp}
\end{equation}
Denoting the order of this approximation by $\epsilon$ and expanding all fields as power series in $\epsilon$, we find that the equations of motion reduce to
\begin{eqnarray}
\epsilon^2 \left[F_0''+\epsilon F_1'' +\mathcal{O}(\epsilon^2)\right] &=& - \left(m^2\,R^2 - \frac{R''}{R} - \Sigma_0'^2\right)F_0  \,\epsilon^2 
\left[1+\left(\frac{F_1}{F_0}-
\frac{2\,\Sigma_0'\,\Sigma_1'}{m^2\,R^2-\Sigma_0'^2- \frac{R''}{R}}\right)\epsilon + \mathcal{O}(\epsilon^2)\right]\,,
\nonumber\\
\epsilon^2 \left[G_0''+\epsilon G_1'' +\mathcal{O}(\epsilon^2)\right]  &=& - \left(\tilde{m}^2\,R^2 - \frac{R''}{R} - \tilde{\Sigma}_0'^2\right)G_0  \,\epsilon^2
\left[1+\left(\frac{G_1}{G_0}-
\frac{2\,\tilde{\Sigma}_0'\,\tilde{\Sigma}_1'}{\tilde{m}^2\,R^2-\tilde{\Sigma}_0'^2- \frac{R''}{R}}\right)\epsilon + \mathcal{O}(\epsilon^2)\right]\,,
\nonumber\\
\epsilon^2 \left[\Sigma_0''+\epsilon \Sigma_1'' +\mathcal{O}(\epsilon^2)\right]  &=& -\frac{2\,F_0'\,\Sigma_0'}{F_0}\epsilon^2 \left[1+\left( -\frac{F_1}{F_0} + \frac{F'_1}{F'_0} + \frac{\Sigma'_1}{\Sigma'_0}\right) \epsilon +\mathcal{O}(\epsilon^2)\right]\,,
\nonumber\\
\epsilon^2 \left[\tilde{\Sigma}_0''+\epsilon \tilde{\Sigma}_1'' +\mathcal{O}(\epsilon^2)\right]  &=& -\frac{2\,G_0'\,\tilde{\Sigma}_0'}{G_0}\epsilon^2 \left[1+\left( -\frac{G_1}{G_0} + \frac{G'_1}{G'_0} + \frac{\tilde{\Sigma}'_1}{\tilde{\Sigma}'_0}\right) \epsilon +\mathcal{O}(\epsilon^2)\right]\,,
\label{ddexp2}
\end{eqnarray}
along with the algebraic relations,
\begin{eqnarray}
\Delta F_1  &=& -\frac{R^2}{F_0}\left(
\frac{\delta m_1^2}{g_1^2} + \frac{\delta m_2^2}{g_2^2}+\frac{\delta m_3^2}{g_3^2}\right)\epsilon^2 
\left(1 -\frac{F_1}{F_0} \epsilon+ \mathcal{O}(\epsilon)^2\right)\,,\nonumber\\
\Delta F_2 &=& \frac{3\,R^2\,\delta m_4^2}{F_0\,g_3^2}\,\epsilon^2 \left(1 -\frac{F_1}{F_0} \epsilon+ \mathcal{O}(\epsilon)^2\right)\,,\nonumber\\
\Delta G_1  &=& \frac{R^2}{G_0}\left(
\frac{\delta m_1^2}{g_1^2} - \frac{2\,\delta m_3^2}{g_3^2}\right)\epsilon^2 
\left(1 -\frac{G_1}{G_0} \epsilon+ \mathcal{O}(\epsilon)^2\right)\,,\nonumber\\
\Delta G_2 &=& \frac{3\,R^2\,\delta m_2^2}{G_0\,g_2^2}\,\epsilon^2 \left(1 -\frac{G_1}{G_0} \epsilon+ \mathcal{O}(\epsilon)^2\right)\,,
\label{hexp2}
\end{eqnarray}
where we defined the mass averages and differences as
\begin{eqnarray}
&\delta m_1^2 \equiv \frac{1}{3} \, \left( 2\,m_1^2- m_2^2 -m_3^2 -2\,m_4^2 +m_5^2+m_6^2\right)\,,&\nonumber\\
&\delta m_2^2 \equiv \frac{1}{3} \, \left( 2\,m_1^2- m_2^2 -m_3^2 -3\,m_5^2+3\,m_6^2\right)\,,&\nonumber\\
&\delta m_3^2 \equiv \frac{1}{9} \, \left( 2\,m_1^2- m_2^2 -m_3^2 +4\,m_4^2 -2\,m_5^2-2\,m_6^2\right)\,,&\nonumber\\
&\delta m_4^2 \equiv -m_2^2 +m_3^2\,,\quad\quad
m^2 \equiv \frac{1}{3} \,\left(m_1^2+m_2^2+m_3^2\right) \,,\quad\quad
\tilde{m}^2 \equiv \frac{1}{3} \,\left(m_4^2+m_5^2+m_6^2\right) \,,&
\end{eqnarray}
The D-terms corresponding to the diagonal generators, at the leading order are then,
\begin{eqnarray}
V_D^{(Y)} &=& \frac{\delta m_1^4}{2\,g_1^2}\,\epsilon^4+\mathcal{O}(\epsilon^6) \,,\nonumber\\
V_D^{(\sigma_3)} &=& \frac{\delta m_2^4}{2\,g_2^2}\,\epsilon^4+\mathcal{O}(\epsilon^6) \,,\nonumber\\
V_D^{(\lambda_3)} &=& \frac{\left(3\,\delta m_3^2 +\delta m_4^2 \right)^2}{8\,g_3^2}\,\epsilon^4+\mathcal{O}(\epsilon^6) \,,\nonumber\\
V_D^{(\lambda_8)} &=& \frac{3\,\left(\delta m_3^2 -\delta m_4^2 \right)^2}{8\,g_3^2}\,\epsilon^4+\mathcal{O}(\epsilon^6) \,,
\end{eqnarray}
As in the two field toy model example, the D-flatness is approximate and of order $\epsilon^4$, and they will dominate over the mass terms much later in the evolution (see the discussion at the end of Section \ref{sec:1fdBG}).

\subsection{Perturbations}\label{sec:2fdPER}
We start by expanding each field as a sum of background and perturbations,
\begin{eqnarray}
\begin{array}{lll}
\phi_1 = \frac{{\rm e}^{i\,\Sigma}}{\sqrt{2}\,R}
  \left(
         \begin{array}{lllll}
         \frac{F_1}{\sqrt{3}}\left[1+ \frac{\delta u_{\bar{1}}^c}{2}\right]\,,
         && \delta u_{\bar{2}}^c\,,
         && \delta u_{\bar{3}}^c
         \end{array}
   \right) \,,
& 
\phi_2 = \frac{{\rm e}^{i\,\Sigma}}{\sqrt{2}\,R}
  \left(
         \begin{array}{lllll}
         \delta s_{\bar{1}}^c\,,
         && \frac{F_2}{\sqrt{3}}\left[1+ \frac{\delta s_{\bar{2}}^c}{2}\right]\,,
         && \delta s_{\bar{3}}^c
         \end{array}
   \right) \,,
\\\\
\phi_3 = \frac{{\rm e}^{i\,\Sigma}}{\sqrt{2}\,R}
  \left(
         \begin{array}{lllll}
         \delta b_{\bar{1}}^c\,,
         && \delta b_{\bar{2}}^c\,,
         && \frac{F_3}{\sqrt{3}}\left[1+ \frac{\delta b_{\bar{3}}^c}{2}\right]
         \end{array}
   \right) \,,
&
\phi_4 = \frac{{\rm e}^{i\,\tilde{\Sigma}}}{\sqrt{2}\,R}
  \left(
         \begin{array}{lllll}
         \frac{G_1}{\sqrt{3}}\left[1+ \frac{\delta d_{\bar{1}}^c}{2}\right]\,,
         && \delta d_{\bar{2}}^c\,,
         && \delta d_{\bar{3}}^c
         \end{array}
   \right) \,,
\\\\
\phi_5 = \frac{{\rm e}^{i\,\tilde{\Sigma}}}{\sqrt{2}\,R}
  \left(
         \begin{array}{l}
         \frac{G_2}{\sqrt{3}}\left[1+\frac{\delta \nu_e}{2}\right]
         \\ \delta e
         \end{array}
   \right) \,,
&
\phi_6 = \frac{{\rm e}^{i\,\tilde{\Sigma}}}{\sqrt{2}\,R}
  \left(
         \begin{array}{lllll}
         \delta c_1 \,,&& \delta c_2 \,,&& \delta c_3 \\
         \frac{G_3}{\sqrt{3}}\left[1+\frac{\delta s_1}{2}\right] \,,&&\delta s_2 \,,&& \delta s_3 
         \end{array}
   \right) \,,
\end{array}
\end{eqnarray}

The perturbations to the field content can be decomposed, in unitary gauge,
\begin{equation}
\begin{array}{lll}
  \delta u_{\bar{1}}^c = \delta f_1 - 2\,\delta f_2 + i \, \delta g_1 \,,&\quad
  \delta u_{\bar{2}}^c = 0\,,&\quad
  \delta u_{\bar{3}}^c = 0 \,, \\
  \delta s_{\bar{1}}^c = -\delta_4 + i\,\delta_1\,,&\quad
  \delta s_{\bar{2}}^c = \delta f_1 + \delta f_2 +\delta f_3 + i \, \delta g_1\,,&\quad
  \delta s_{\bar{3}}^c = \frac{F_2}{\sqrt{F_2^2+F_3^2}}\left(\delta_{14} + i\,\delta_{13}\right)  \,,\\
  \delta b_{\bar{1}}^c = -\delta_{10} + i\,\delta_7 \,,&\quad
  \delta b_{\bar{2}}^c = \frac{F_3}{\sqrt{F_2^2+F_3^2}}\left(\delta_{14} - i\,\delta_{13}  \right)\,,&\quad
  \delta b_{\bar{3}}^c = \delta f_1 + \delta f_2 -\delta f_3 + i \, \delta g_1\,,\\
  \delta d_{\bar{1}}^c = \delta f_4 - 2\,\delta f_5 + i \, \delta g_2 \,,&\quad
  \delta d_{\bar{2}}^c = \delta_{6} + i\,\delta_{3}  \,,&\quad
  \delta d_{\bar{3}}^c = \delta_{12} + i\,\delta_9 \,,\\
  \delta \nu_e = \delta f_4 + \delta f_5 + \delta f_6 + i \, \delta g_2 \,,&\quad
  \delta e = \frac{G_2}{\sqrt{G_2^2+G_3^2}}\left(\delta_{16} - i \, \delta_{15}\right)\,, &\\  
  \delta c_1 = \frac{G_3}{\sqrt{G_2^2+G_3^2}}\left(\delta_{16} + i \, \delta_{15} \right)\,,&\quad
  \delta c_2 = \delta_{17} + i \, \delta_{18}   \,,&\quad
  \delta c_3 = \delta_{19} + i \, \delta_{20} \,,\\
  \delta s_1 = \delta f_4 + \delta f_5 - \delta f_6 + i \, \delta g_2 \,,&\quad
  \delta s_2 = -\delta_{5} + i \, \delta_{2}   \,,&\quad
  \delta s_3 = -\delta_{11} + i \, \delta_{8} \,,
\end{array}
\end{equation}
This system, including the gauge fields, has $64$ degrees of freedom. From the 3-dimensional rotational symmetry of the Lagrangian, we know that all the transverse vector degrees ($24$) decouple from the remaining $40$ scalar degrees. Furthermore, since all the gauge symmetries are broken, the transverse vector modes will acquire masses of order VEV, suppressing the non-perturbative effects. The quadratic action of the perturbations of the model can be decomposed into $9$ decoupled subsystems, which are studied in detail in Appendix \ref{AppD}. Formally, we have, 
\begin{equation}
S^{(2)} = S_\perp^{\rm non~diag} + S_\perp^{\rm diag} + S_{c_2}+ S_{c_3} + S_{e,\,c_1}+ S_{b_{\bar{2}} ,\,s_{\bar{3}}} + S_{\delta f,\,\delta g}  +S_{s_{\bar{1}} ,\,s_2,\,d_{\bar{2}}} + S_{b_{\bar{1}} ,\,s_3,\,d_{\bar{3}}}\,,
\end{equation}
Next, we diagonalize each of these subsystems and determine the physical modes. Due to the complexity of the model, the diagonalization procedure needs to be carried out in TeV/VEV expansion. In Table \ref{tab-deg}, we summarize the results of the detailed study of the spectrum. The heavy eigenstates which have the same mass as the transverse vectors in the accuracy of the expansion are identified as ``Longitudinal vector modes'', whereas the remaining heavy modes with additional terms in the sub-leading order are labeled ``Higgs''. In this example, we also encounter ``other heavy'' modes, which have masses of the order of VEV, although these masses are provided by the F-terms, and are proportional to the Yukawa coupling $y_d$. Additionally, we have $2$ light degrees of freedom per flat direction, corresponding to their fluctuations. All the remaining degrees of freedom are named ``Other light''. 
\begin{table*}
\begin{tabular}{|c||c|c|c|c|c|c||c|}
\hline
&$\perp$ Vector
&$\parallel$ Vector  
&Higgs
&Flat Dir.
& Other heavy
&Other light
&Total\\
\hline \hline
$S_\perp^{\rm non~diag}$ &$16$& --&--&--&--&--&$16$\\\hline
$S_\perp^{\rm diag}$ &$8$& --&--&--&--&--&$8$\\\hline
$S_{c_2}$&--& --&--&--&$2$&--&$2$\\\hline
$S_{c_3}$&--& --&--&--&--&$2$&$2$\\\hline
$S_{e,\,c_1}$&--& $2$&$2$&--&--&--&$4$\\\hline
$S_{b_{\bar{2}} ,\,s_{\bar{3}}}$&--& $2$&$2$&--&--&--&$4$\\\hline
$S_{\delta f,\,\delta g}$&--& $4$&$4$&$4$&--&--&$12$\\\hline
$S_{s_{\bar{1}} ,\,s_2,\,d_{\bar{2}}}$&--& $2$&$2$&--&$2$&$2$&$8$\\\hline
$S_{b_{\bar{1}} ,\,s_3,\,d_{\bar{3}}}$&--& $2$&$2$&--&--&$4$&$8$\\\hline\hline
Total & $24$ & $12$ & $12$ & $4$ & $4$& $8$&$64$\\\hline\hline
\end{tabular}%
\caption{Table summarizing the classification of degrees of freedom for the decoupled sub-systems. }
\label{tab-deg}
\end{table*}

The first seven subsystems do not have the basic ingredients for the non-perturbative decay. However, the last two subsystems have rather non-trivial mixings and they indeed provide the non-adiabatic rotation of the eigenstates. The system $S_{s_{\bar{1}} ,\,s_2,\,d_{\bar{2}}}$ which consists of the perturbations of $s_{\bar{1}}^c$, $s_2$ and $d_{\bar{2}}^c$ coupled to the longitudinal vector modes $A^{(\lambda_1)}_L$ and $A^{(\lambda_2)}_L$, contains a pair of light particles and three pairs of heavy particles. We find that, in the $\epsilon$ expansion, the  (heavy) longitudinal vectors decouple and we get $\mathcal{O}(\epsilon^0)$ adiabaticity matrix components corresponding to mixings among the light modes and the Higgses. In other words, this system has non-adiabatic rotation which may give rise to a non-perturbative decay\footnote{We note that for the system $S_{s_{\bar{1}} ,\,s_2,\,d_{\bar{2}}}$, the $\Gamma$ matrix is independent of the mass differences and do not undergo any simplification even in the degenerate mass limit.}. However, to quantify this, a numerical study is required. In such a study, one still needs to include components of the $\Gamma$ matrix which do not cause production, but are responsible in the rotation of the produced quanta into other states (eg. mixings between the Higgs and the other heavy particles). This is beyond the scope of the present paper.

On the other hand, the final subsystem can be simplified down to a more familiar problem in some limits, and we can extract a clearer picture and a more definite result out of it. This system, consisting of perturbations to $b_{\bar{1}}^c$, $s_3$ and $d_{\bar{3}}^c$ coupled to the longitudinal vector modes $A^{(\lambda_4)}_L$ and $A^{(\lambda_5)}_L$,  contains two pairs of light particles and two pairs of heavy particles. Upon calculation of the $\Gamma$ matrix in $\epsilon$ expansion, the longitudinal vector modes can be decoupled as before, leaving a coupled system of six degrees of freedom with order $\epsilon^0$ adiabaticity matrix elements. In other words, at this level, this system has more room for production than the previous system, due to the doubling of the light modes. The problem can be further simplified by tuning the mass parameters. In the limit of mass degeneracy $m_1=m_2=m_3=m$, $m_4=m_5=m_6=\tilde{m}$, we see that both the eigenvalues and the $\Gamma$ matrix components produce two exact copies of the coupled system found in the four field toy model of \cite{Gumrukcuoglu:2008fk}, with $g_3^2 \rightarrow 3\,e^2/2$. That is, we end up with two identical systems, each consisting of two light modes and a Higgs, where particle production is due to the mixing between the two light modes and between the Higgs and one of the light modes, as well as the non-adiabatic evolution of the $\mathcal{O}(\epsilon)$ frequencies. A less strict tuning, introduced by having
\begin{equation}
\Delta M \equiv \frac{2}{9}  \,\frac{\left(m_1^2+m_2^2-2\,m_3^2+2\,m_4^2-m_5^2-m_6^2 \right)R^2\,\left(F_0^2+G_0^2\right)^2}{\left(G_0 F_0'-F_0 G_0'\right)^2+F_0^2G_0^2 \left(\Sigma_0'-\tilde{\Sigma}_0'\right)^2} \ll 1\,,
\label{deltamcom}
\end{equation}
gives exactly the same $\Gamma$ matrix elements at order $\Delta M^0$. The only differences are the other combinations of mass differences in the sub-leading terms of the $\mathcal{O}(\epsilon^0)$ frequencies, as well as the high order contributions to the VEVs, which are non-zero in the absence of complete degeneracy. These sub-leading terms however have negligible effect on the occupation numbers. Therefore, for the case where $\Delta M \ll 1$, even the non-degenerate system can be described by the results of the four field toy model.

The function $\Delta M$ is time dependent and in the course of the evolution, it oscillates. It is clear that even if $\Delta M \ll 1$ is satisfied at one time, it may be violated at some other one. However, we can estimate a condition on only the mass parameters by approximating $F_0' \sim m\,R\,F_0$, $G_0' \sim \tilde{m}\,R\,G_0$, $\Sigma_0' \sim m\,R$, $\tilde{\Sigma}_0' \sim \tilde{m}\,R$. Furthermore, since the production occurs when the instantaneous VEVs are comparable \cite{Allahverdi:2006xh,Gumrukcuoglu:2008fk}, we also take $G_0\sim F_0$ in order to guarantee that the condition $\Delta M \ll 1$ is satisfied when production is expected. With these considerations, (\ref{deltamcom}) reduces to a time independent condition involving only the mass parameters
\begin{equation}
\frac{3\,\delta m_3^2-\delta m_4^2}{\left(m-\tilde{m}\right)^2} \ll 1\,.
\end{equation}
As long as this condition applies, the results of \cite{Gumrukcuoglu:2008fk} show that the flat directions decay within the $\mathcal{O}(10)$ rotations, for a range of three orders of magnitude in the ratio of the initial VEVs (see Figure 4 of \cite{Gumrukcuoglu:2008fk}). If this condition is not satisfied, we still expect a non-perturbative decay; however, to quantify it requires an extensive numerical evolution.


\section{Discussion}\label{sec:disc}

The primary aim of this paper was to see if the extensive numerical study carried out in the framework of $4$-field gauged toy model \cite{Gumrukcuoglu:2008fk} can accurately describe a realistic case. We have shown that this toy model, which has only $U(1)$ symmetry, two mass parameters and no F-terms, provides a very good description of an example from MSSM, where both $u^cd^cd^c$ and $Q L d^c$ flat directions are present simultaneously. Specifically, both cases contain ``independent flat directions'', that is, the fields in one flat direction are decoupled from the others at the background level. As a result, the background equations of motion for the two models have the same form. Furthermore, we found that out of the $64$ real degrees of freedom in the problem, only the two decoupled actions $S_{s_{\bar{1}},s_2, d_{\bar{2}}}$ and $S_{b_{\bar{1}},s_3, d_{\bar{3}}}$, each consisting of 8 degrees, exhibit the non-adiabatic evolution of the eigenstates. At the limit of degenerate masses, that is, when the masses of the fields in a flat direction are identical, we found that $S_{b_{\bar{1}},s_3, d_{\bar{3}}}$ can be decomposed into two copies of the coupled system in the $4$ field gauged toy model. Therefore, it is safe to state that, in the degenerate mass case, the numerical results of the toy model provides an exact description of this part of the action. If the action $S_{s_{\bar{1}},s_2, d_{\bar{2}}}$ contributes to the non-perturbative decay as we expect, the resulting quanta will be different ones and their production will not effect the occupation numbers of generated $b_{\bar{1}}$, $s_3$ and $d_{\bar{3}}$ perturbations, as long as the linearized approximation holds. 

Another focus of the present paper was to understand the effect of different soft supersymmetry breaking masses to the non-perturbative production. For an exactly D-flat direction, the fields in that direction have the same soft mass. Conversely, if all the fields have different masses, the D-term will be non-zero and proportional to the fourth power of mass differences. As a reference calculation, we showed that for a single flat direction, the flat direction perturbations still decouple at the leading order, thus verifying that there is no non-perturbative decay in this case. The effect of different masses in the $u^cd^cd^c + Q L d^c$ example had similar consequences for the flat direction perturbations. Furthermore, the four field toy model with degenerate masses still describes the subsystem $S_{b_{\bar{1}},s_3, d_{\bar{3}}}$ exactly when the condition (\ref{deltamcom}) holds. Therefore, based on the results of \cite{Gumrukcuoglu:2008fk}, we conclude that the flat directions $u^cd^cd^c + Q L d^c$ decay in $\mathcal{O}(10)$ rotations, also for this case. 

One issue about the example considered here is the hierarchy between the VEVs corresponding to each flat direction. The first non-renormalizable operator present for $u^cd^cd^c$ flat direction has $d=6$, whereas for $Q L d^c$, it has $d=4$ \cite{Gherghetta:1995dv}, resulting in an initial VEV ratio of $\sim10^3$. This ratio, although large, may still give rise to a production if the orbits have enough ellipticity and the mass ratio is large enough. For instance, in the numerical results of \cite{Gumrukcuoglu:2008fk}, it was shown that for two flat directions with mass ratio $\tilde{m}/m =7.63$, they decay within $20$ rotations for initial VEV ratio of $10^3$. If the orbits are closer to the radial one, one might still have a non-perturbative decay with a smaller mass ratio for the same VEV hierarchy. On the other hand, the non-renormalizable terms are model dependent, and they may be forbidden by some discrete symmetries. In fact, to recover the conditions of delayed thermalization \cite{Allahverdi:2005mz}, one needs $\phi_{\rm in} \gtrsim 10^{-2} M_p$, which requires all terms with $d < 11$ to vanish \cite{Olive:2006uw}.

Although the numerical results of \cite{Gumrukcuoglu:2008fk} shows that the flat directions may decay non-perturbatively, it is still not clear how this effect changes the picture of thermalization. Specifically, the new quanta are produced in a resonant band with momenta $k \lesssim m$ which is still non-relativistic. Since the variances are large, the gauge fields will still have large enough mass contributions to suppress the scatterings between inflaton decay products. On the other hand, tracking the evolution of the produced particle distribution is beyond the reach of the linearized calculation. One needs to control the back reaction effects to determine how fast the variances decrease. In such a computation, the distribution is likely to thermalize, possibly not in $\mathcal{O}(10)$ rotations, but we expect it to be much earlier than $10^{11}$ rotations that is required by the perturbative decay \footnote{Indeed, our preliminary results from lattice simulations of non-gauged toy model \cite{Olive:2006uw} up to $\mathcal{O}(100)$ rotations show the beginning of thermalization of the produced particles, i.e. the initial distribution with momenta $k \lesssim m$ starts to extend toward high momentum region.}.

It is clear that the linear study done in the present paper, along with \cite{Olive:2006uw, Basboll:2007vt, Basboll:2008gc, Gumrukcuoglu:2008fk} are limited to the stages of the evolution until the production is significant, and they only provide a glimpse to the beginning of the non-perturbative decay. For instance, we found that the $u^cd^cd^c + Q L d^c$ example should result in a decay within $\mathcal{O}(10)$ rotations for the range of parameters in \cite{Gumrukcuoglu:2008fk}. However, once non-linear effects become important, the particles produced through different decoupled actions may interact and change the outcome of this study. Therefore, one should be cautious to extrapolate a non-linear study based on the toy model to a realistic one. 

The logical step to be taken next is to include higher order terms and study the effect on the decay time and thermalization of the produced quanta. In a recent study \cite{Dufaux:2009wn}, the non-gauged toy model of \cite{Olive:2006uw} was evolved on a lattice using the ClusterEasy code \cite{Felder:2000hq}, verifying that the non-perturbative decay is still realized within the first few rotations. Another interesting result of \cite{Dufaux:2009wn} was the calculation of the gravity waves, sourced by the quick decay of the flat directions. Their spectrum was found to fall naturally into Hz-kHz range and depending on the initial VEV of the flat directions, may potentially be within the reach of upcoming experiments, such as Advanced LIGO. For more realistic models with gauge fields, the resulting spectrum may be different, but since the mass scale of the flat directions is of order TeV, it will have a frequency range similar to that in the toy model. In a future work, we will address the effect of back-reaction and gravity wave production in the framework of gauged models.

However, there are still some problems left that can be dealt with analytical tools. In \cite{CyrRacine:2009qk}, it was shown that the non-perturbative decay of flat directions may have an observable effect through the amplification of curvature perturbations. In the context of natural supergravity inflation \cite{Kawasaki:2000ws}, Ref. \cite{Kaminska:2009wh} showed that the non-perturbative decay of flat directions allows the inflaton preheating to be realized in these models. On the other hand, even at the linearized level, we do not have a complete numerical study of the decay for ``overlapping flat directions'', although we have approximate calculations showing the non-adiabatic rotation occurs \cite{Basboll:2007vt, Basboll:2008gc}. Specifically, the toy model of \cite{Basboll:2007vt} is qualitatively different than the ones considered in the present work, as well as \cite{Gumrukcuoglu:2008fk}, in the sense that the flat directions are coupled at the background level. A numerical analysis in the fashion of \cite{Gumrukcuoglu:2008fk} would be very useful in understanding the time scale of the decay and the range of hierarchy that allows production. There are many examples with ``overlapping flat directions'' in MSSM which provide the necessary ingredients for non-adiabatic evolution (e.g. \cite{Basboll:2008gc}), and it is an interesting challenge to find simple toy models that correctly describe at least parts of these examples.

\begin{acknowledgments}
I would like to thank A. Basb\o{}ll, J. F. Dufaux, K. A. Olive and M. Sexton for useful discussions. I am also grateful to M. Peloso for stimulating discussions and valuable comments on the manuscript. This work was partially supported by the DOE grant DE-FG02-94ER-40823.
\end{acknowledgments}

\appendix


\section{Simplifying a coupled system}\label{AppA}

In this section, we summarize the steps for transforming any given action to the form (\ref{canform}) and conclude by discussing the order in the TeV/VEV expansion (\ref{2fdapp}) that we need to calculate. We start from an action of $N$ coupled scalars
\begin{eqnarray}
S= \int d^3k\,d\eta \,\mathcal{L}_k\,,
\end{eqnarray}
where the Lagrangian density in Fourier space $\mathcal{L}_k$ is, generically
\begin{equation}
\mathcal{L}_k=\frac{1}{2} \left(\phi^{\prime\,\dagger} \,T_1\,\phi'+\phi^{\prime \, \dagger}\,K_1\,\phi +\phi^{\dagger}\,K_1^T\,\phi' -\phi^\dagger\,\Omega^2_1\,\phi\right)\,.
\label{initform}
\end{equation}
$T_1$ and $\Omega^2_1$ are symmetric and time dependent $N \times N$ matrices, whereas the real matrix $K_1$ is in general asymmetric. Due to the isotropy of the problem, these matrices are invariant under the reflection operation $k\rightarrow -k$. We now proceed to simplify the form of this action to that of coupled oscillators in Minkowski background. We first diagonalize the kinetic terms with orthogonal matrix $R_1$
\begin{equation}
R_1^T\, T_1\,R_1 =T_2 \;{\rm (diagonal)}\,.
\end{equation}
We then write the Lagrangian in the new basis $\psi \equiv R_1^T \phi$ as
\begin{equation}
\mathcal{L}_k=\frac{1}{2} \left(\psi^{\prime\,\dagger} \,T_2\,\psi'+\psi^{\prime \, \dagger}\,K_2\,\psi +\psi^{\dagger}\,K_2^T\,\psi' -\psi^\dagger\,\Omega^2_2\,\psi\right)\,,
\end{equation}
with
\begin{equation}
K_2 \equiv R_1^T\,K_1\,R_1+R_1^T\,T_1\,R_1' \,,\quad\quad
\Omega_2^2 \equiv R_1^T\,\Omega^2_1\,R_1-R_1'^T\,T_1\,R_1'-2\,R_1'^T\,K_1\,R_1\,.
\end{equation}
Next, we rescale the fields such that the kinetic matrix becomes unity
\begin{equation}
R_2 \equiv \left(T_2\right)^{-1/2} = R_2^T \,.
\end{equation}
In the rescaled basis $\chi \equiv R_2^{-1} \psi$, the Lagrangian reads
\begin{equation}
\mathcal{L}_k=\frac{1}{2} \left(\chi^{\prime\,\dagger} \chi'+\chi^{\prime \, \dagger}\,K_3\,\chi+\chi^{\dagger}\,K_3^T\,\chi' -\chi^\dagger\,\Omega^2_3\,\chi\right)\,,
\end{equation}
with
\begin{equation}
K_3 \equiv R_2\,K_2\,R_2 + R_2 \,T_2\,R_2'\,,\quad\quad
\Omega^2_3 \equiv R_2\,\Omega^2_2\,R_2  - R_2'\,T_2\,R_2' - 2\,R_2'\,K_2\,R_2 \,.
\end{equation}
By adding to the Lagrangian, the boundary term
\begin{equation}
 \mathcal{L}_k \rightarrow \mathcal{L}_k -\frac{1}{4}\,\frac{d}{d\eta} \left[ \chi^\dagger \left(K_3+K_3^T\right)\chi\right]\,,
\end{equation}
we obtain a more convenient form, where the matrix which mixes the fields to their derivatives is anti-symmetric
\begin{equation}
\mathcal{L}_k=\frac{1}{2} \left(\chi^{\prime\,\dagger} \chi'+\chi^{\prime \, \dagger}\,K_4\,\chi-\chi^{\dagger}\,K_4\,\chi' -\chi^\dagger\,\Omega^2_4\,\chi\right)\,,
\end{equation}
where
\begin{equation}
K_4 = -K_4^T \equiv \frac{1}{2} \left(K_3-K_3^T\right)\,,\quad\quad
\Omega^2_4 \equiv \Omega^2_3+\frac{1}{2}\left(K_3+K_3^T\right)'\,.
\end{equation}
Finally, to remove the mixing matrix $K_4$, we do a further transformation with $\Psi\equiv R_5\, \chi$, satisfying $R_5^T R_5'=K_4$. With this rotation, we obtain an action resembling coupled oscillators in Minkowski space (\ref{canform})
\begin{equation}
S = \frac{1}{2}\int d^3k\,d\eta\left[\Psi^{\prime \,\dagger} \Psi' - \Psi^\dagger R_5 \left(\Omega^2_4+K_4^T K_4\right)R_5^T \Psi\right]\,.
\label{canform2}
\end{equation}
As outlined in \cite{Gumrukcuoglu:2008fk}, the construction of the matrix $\Gamma=C^T\,C'$ that is needed for the computation of evolution equations, does not require the explicit knowledge of $R_5$. Diagonalizing simply the matrix $\left(\Omega_4^2+K_4^TK_4\right) $, through
\begin{equation}
\xi^T \left(\Omega_4^2+K_4^TK_4\right) \xi = \omega^2\;{\rm (diagonal)}\,,
\end{equation}
one finds the relation $C = R_5 \,\xi$ which gives
\begin{equation}
\Gamma = \xi^T \xi' + \xi^T K_4 \xi\,.
\label{gamfin}
\end{equation}
For the example in Section \ref{sec:2fd}, the action is very complicated and even the first steps where the kinetic matrix $T_1$ is diagonalized cannot be done exactly. Therefore, for some of the subsystems we encounter, we need to apply the expansion (\ref{2fdapp}) at the level of eq (\ref{initform}). In Section \ref{sec:formalism}, we have shown that the matrix $\Gamma$ is needed at the order $\epsilon$. Going backwards from (\ref{gamfin}), we see that to calculate this matrix at the required order, the matrix $\xi$ is needed at order $\epsilon^0$, all the $K_i$ matrices at order $\epsilon$, and matrices $R_1$, $R_2$, $T_i$, $\Omega_i^2$ and $\omega^2$ are needed at order $\epsilon^2$. 

\section{Calculations for 2 field model} \label{AppB}

In this section, we summarize the calculation for the perturbations in $2$ field model of Section \ref{sec:1fdPER}.
The quadratic action for the transverse vector modes is immediately decoupled:
\begin{equation}
S_\perp = \frac{1}{2} \int d^4x \left[A_i^{T \prime} A_i^{T \prime} -(\partial_i A_j^T) (\partial_i A_j^T) - \frac{e^2\,(F^2+\Delta F^2)}{4}A_i^T A_i^T\right] \,.
\end{equation}
The remaining action consists of a system of four coupled real fields,
\begin{eqnarray}
S_{\rm 4 \,dof} &=& \frac{1}{2}\int d^4 x \Bigg\{
\delta_H'^2-(\partial_i \delta_H)^2-\mu^2_{\delta_H} \delta_H^2
+\left(\partial_i L'\right)^2 -\frac{e^2}{4}\left(F^2+ \Delta F^2\right)\,\left(\partial_i L\right)^2
\nonumber\\ 
&&\quad\quad\quad\quad
+\left(\partial_i A_0\right)^2 -2\,\left(\partial_i L'\right)\left(\partial_i A_0\right) +\frac{e^2}{4} \left(F^2+ \Delta F^2\right)A_0^2
\nonumber\\ 
&&\quad\quad\quad\quad
+r'^2-(\partial_i r)^2 -\mu_r^2\,r^2+\left(F^2+ \Delta F^2\right)\left[ \sigma'^2-(\partial_i \sigma)^2 \right]-\mu^2_\sigma \sigma^2
\nonumber\\ 
&&\quad\quad\quad\quad
-2\,\mu^2_{r\,\delta_H}\,r \,\delta_H
+4\,\lambda\left(F^2-\Delta F^2\right)\sin(4\,\Sigma)\,\left(F\,r-\Delta F\,\delta_H\right)\,\sigma
\nonumber\\ 
&&\quad\quad\quad\quad
+\frac{2\,e\left(F^2-\Delta F^2\right)\,\Sigma'}{F^2+\Delta F^2}\left(\Delta F\,r-F\,\delta_H\right)A_0
+2\,e\,F\,\Delta F\,(\partial_i L)(\partial_i \sigma)
\nonumber\\ 
&&\quad\quad\quad\quad
+\left[\frac{4\,\left(F^2-\Delta F^2\right)\,\Sigma'}{F^2+\Delta F^2}\,\left(F\,r-\Delta F\,\delta_H\right)-2\,e\,F\,\Delta F\,A_0 \right] \sigma' \Bigg\}
\label{act1fdpert}
\end{eqnarray}
where,
\begin{eqnarray}
\mu^2_{\delta_H} &\equiv& \frac{e^2\,F^2}{4} +m^2\,R^2 -\frac{R''}{R} -\left(1+\frac{4\,F^2\,\Delta F^2}{\left(F^2+\Delta F^2\right)^2}\right)\Sigma'^2 	-\frac{\lambda}{2}\,(F^2-3\,\Delta F^2)\,\cos(4\,\Sigma)
\,,\nonumber\\
\mu^2_r &\equiv& \frac{e^2\,\Delta F^2}{4} +m^2\,R^2 -\frac{R''}{R}-			\left(1+\frac{4\,F^2\,\Delta F^2}{\left(F^2+\Delta F^2\right)^2}\right)\Sigma'^2-\frac{\lambda}{2}\,(\Delta F^2-3\,F^2)\,\cos(4\,\Sigma)
\,,\nonumber\\
\mu^2_\sigma &\equiv & -2\,\lambda\,\left(F^2-\Delta F^2\right)^2\,\cos(4\,\Sigma)
\,,\nonumber\\
\mu^2_{r\,\delta_H}&\equiv & F\,\Delta F\left(\frac{e^2}{2}-\lambda\,\cos(4\,\Sigma) +\frac{4\,\Sigma'^2}{F^2+\Delta F^2}\right) + \delta m^2\,R^2 \,.
\end{eqnarray}
Contrary to the degenerate mass case ($\delta m =0$, $\Delta F=0$), $r$ and $\sigma$, which were the perturbations of flat directions, are now coupled to the Higgs and the longitudinal vector through the non-zero VEV difference $\langle \phi_1\rangle-\langle\phi_2\rangle$.

We then expand the fields in terms of plain waves, and solve the constraint equation for the non-dynamical degree $A_0$, which yields
\begin{equation}
A_0=\left(1+\frac{e^2\left(F^2+\Delta F^2\right)}{4\,k^2}\right)^{-1}
\left[L'+\frac{e\,F\,\Delta F}{k^2}\,\sigma' -\frac{e\,\left(F^2-\Delta F^2\right)\Sigma'}{k^2\left(F^2+\Delta F^2\right)} \left(\Delta F\,r-F\,\delta_H\right) \right]\,.
\label{intoutA0}
\end{equation}
Using the expression above in (\ref{act1fdpert}), the action becomes of the form (\ref{initform}). By a series of redefinitions as described in Appendix \ref{AppA}, we obtain the form (\ref{canform}). The explicit expressions for the matrices $K_4$ and $\Omega^2_4$ are too involved for presentation. However, one can write them as an expansion series in using the $\epsilon$ expansion (\ref{1fdapp}) and background expressions (\ref{ddexp}) and (\ref{hexp}). The non-zero components of $K_4$ at $\mathcal{O}(\epsilon)$ are
\begin{equation}
\left[K_4\right]_{13}  = \left[K_4\right]_{14}  =  \left\{-\frac{\Sigma_0'}{\sqrt{2}}\right\}\,,
\end{equation}
with non-zero $\Omega_4^2$ components up to $\mathcal{O}(\epsilon^2)$,
\begin{eqnarray}
\left[\Omega_4\right]_{11} &=& \left[ k^2+m^2\,R^2-\Sigma_0'^2 - \frac{R''}{R}+ \frac{3\,\lambda}{2}\,F_0^2\,\cos(4\,\Sigma_0)\right] \,,\nonumber\\
\left[\Omega_4\right]_{12} &=&\left[-\delta m^2\,R^2\right] 
\,,\nonumber\\
\left[\Omega_4\right]_{13} &=&\left[\Omega_4\right]_{14} = \left[-\frac{3\,\lambda\,F_0^2\,\sin(4\,\Sigma_0)}{2\,\sqrt{2}} \right]\,,\nonumber\\
\left[\Omega_4\right]_{22} &=& \frac{e^2}{4}\,F_0^2 + \left\{ \frac{e^2}{2}\,F_0\,F_1\right\} \nonumber\\
&&+ \left[k^2 +m^2 \,R^2 +3\,\Sigma_0'^2 +\frac{e^2}{4}\left(F_1^2+2 \,F_0\,F_2\right) -\frac{R''}{R}-\frac{\lambda}{2}\,F_0^2\,\cos(4\,\Sigma_0)\right]
\,,\nonumber\\
\left[\Omega_4\right]_{33} &=& \left[\Omega_4\right]_{44}  = \frac{e^2}{8}\,F_0^2 + \left\{ \frac{e^2}{4}\,F_0\,F_1\right\} \nonumber\\
&&\quad\quad\quad\quad+ \left[k^2 +\frac{m^2}{2} \,R^2 -\frac{1}{2}\Sigma_0'^2 +\frac{e^2}{8}\left(F_1^2+2 \,F_0\,F_2\right) -\frac{R''}{2\,R}-\frac{3\,\lambda}{4}\,F_0^2\,\cos(4\,\Sigma_0)\right] 
\,,\nonumber\\
\left[\Omega_4\right]_{34} &=& -\frac{e^2}{8}\,F_0^2 - \left\{ \frac{e^2}{4}\,F_0\,F_1\right\} \nonumber\\
&&+ \left[\frac{m^2}{2} \,R^2 -\frac{1}{2}\Sigma_0'^2 -\frac{e^2}{8}\left(F_1^2+2 \,F_0\,F_2\right) -\frac{R''}{2\,R}-\frac{3\,\lambda}{4}\,F_0^2\,\cos(4\,\Sigma_0)\right] 
\,,
\end{eqnarray}
In the above expressions and elsewhere, terms outside parenthesis, within curly brackets and square brackets are, respectively, of zeroth, first and second order in $\epsilon$.

To determine the adiabaticity conditions, the only remaining step is to diagonalize the $\Omega_4^2+K_4^T\,K_4$ matrix. The eigenfrequencies of the physical modes are
\begin{eqnarray}
\omega_A^2 &=& \left[k^2+m^2\,R^2-\frac{3}{2}\,\lambda\,F_0^2-\frac{R''}{R}\right]
\,,\nonumber\\
\omega_B^2 &=& \left[k^2+m^2\,R^2+\frac{3}{2}\,\lambda\,F_0^2-\frac{R''}{R}\right]
\,,\nonumber\\
\omega_C^2 &=& \frac{e^2}{4}\,F_0^2 + \left\{ \frac{e^2}{2}\,F_0\,F_1\right\} +\left[k^2 +\frac{e^2}{4}\left(F_1^2+2\,F_0\,F_2\right)\right]
\,,\nonumber\\
\omega_D^2 &=& \frac{e^2}{4}\,F_0^2 + \left\{ \frac{e^2}{2}\,F_0\,F_1\right\} +\left[k^2 +\frac{e^2}{4}\left(F_1^2+2\,F_0\,F_2\right)+m^2\,R^2+3\,\Sigma_0'^2\right]\,.
\end{eqnarray}
In the first two lines, we have kept the sub-leading terms proportional to $\lambda$ and $R''$ for comparison with the degenerate mass case. At the given order of expansion, the two modes coincide with the perturbations to the flat direction in \cite{Gumrukcuoglu:2008fk}. The last two lines show the frequencies corresponding to the longitudinal component of the $U(1)$ gauge field and the physical Higgs, respectively. The transformation matrix $\xi$ which diagonalizes $\Omega_4^2+K_4^T\,K_4$ matrix is
\begin{equation}
\xi = \frac{1}{\sqrt{2}}\left(
\begin{array}{llll}
\sqrt{2} \,\sin(2\,\Sigma_0) & -\sqrt{2}\,\cos(2\,\Sigma_0) & 0 & 0\\
0 & 0 & 0 & \sqrt{2}\\
\cos(2\,\Sigma_0) & \sin(2\,\Sigma_0) &-1  & 0 \\   
\cos(2\,\Sigma_0) & \sin(2\,\Sigma_0) &1  & 0   
\end{array}
\right)
+\mathcal{O}(\epsilon)
\end{equation}
Using this, we compute the $\Gamma$ matrix at order $\epsilon$,
\begin{equation}
\Gamma = \epsilon\left(
\begin{array}{llll}
0 & \Sigma_0' & 0 & 0 \\
-\Sigma_0' & 0& 0 & 0 \\
0 & 0 & 0 & 0\\
0 & 0 & 0 & 0   
\end{array}
\right)
+\mathcal{O}(\epsilon^2)\,.
\end{equation}
Therefore, the only mixing is between the two light modes. However, shortly after the VEV starts oscillating, the two light modes become degenerate due to the suppression of the term proportional to $\lambda$. Based on the arguments in Section \ref{sec:formalism}, the leading terms in the adiabaticity matrix $\mathcal{A}$ (\ref{adia}) are then at least of order $\epsilon$. Just like in the degenerate mass case, this model has no non-perturbative decay channel. Additionally, due to lack of mixing in the $\Gamma$ matrix between the flat direction perturbations and other modes, we conclude that the flat direction perturbations decouple from the rest of the system at the given order in $\epsilon$ expansion.


\section{Background calculations for the $u^c d^c d^c + Q L d^c$ example}\label{AppC}
In this appendix, we summarize the calculations for the background quantities in the realistic 2 flat direction example in Section \ref{sec:2fdBG}. For clearer notation, we adopt the following decomposition:
\begin{eqnarray}
&&\Phi_1 = \frac{ F_1}{\sqrt{6}\,R}\,{\rm e}^{i\,\Sigma_1}\,, \quad
\Phi_2 = \frac{ F_2}{\sqrt{6}\,R}\,{\rm e}^{i\,\Sigma_2}\,, \quad
\Phi_3 = \frac{ F_3}{\sqrt{6}\,R}\,{\rm e}^{i\,\Sigma_3}\,, \nonumber\\
&&\Phi_4 = \frac{ G_1}{\sqrt{6}\,R}\,{\rm e}^{i\,\tilde{\Sigma}_1}\,, \quad
\Phi_5 = \frac{ G_2}{\sqrt{6}\,R}\,{\rm e}^{i\,\tilde{\Sigma}_2}\,, \quad
\Phi_6 = \frac{ G_3}{\sqrt{6}\,R}\,{\rm e}^{i\,\tilde{\Sigma}_3}\,, 
\end{eqnarray}
Without fixing the gauge freedom, we write the background action as
\begin{eqnarray}
S_{\rm bg} &=& \frac{1}{2}\int dx^4 \Bigg\{\frac{1}{3} \sum_{i=1}^3 \left[ F_i^{\prime\,2} + G_i^{\prime\,2}  
- \left(m_i^2\,R^2 - \frac{R''}{R} -\Sigma_i^{\prime\,2} \right)F_i^2 - \left(m_{i+3}^2\,R^2 - \frac{R''}{R} -\tilde{\Sigma}_i^{\prime\,2} \right)G_i^2 \right] 
\nonumber\\&&\quad\quad\quad\quad
+\frac{g_1^2}{1296}\left(4\,F_1^2 - 2\,F_2^2 -2\, F_3^2-2\,G_1^2+3\,G_2^2 -G_3^2\right)^2 
- \frac{g_2^2}{144}\,\left(G_2^2-G_3^2\right)^2 
\nonumber\\&&\quad\quad\quad\quad
-\frac{g_3^2}{432} \left[ \left(F_1^2 + F_2^2 -2\,F_3^2 +G_1^2 - G_3^2\right)^2 +3\, \left(F_1^2 -F_2^2 +G_1^2 -G_3^2\right)^2\right]
\nonumber\\&&\quad\quad\quad\quad
+\frac{g_1^2}{108}\,A_0^{(Y)^2}\left(16\,F_1^2 + 4\,F_2^2 + 4\,F_3^2 + 4\, G_1^2 + 9\,G_2^2 +G_3^2\right)
+\frac{g_2^2}{12} \, \left(A_0^{(\sigma_1)^2}+A_0^{(\sigma_2)^2}+A_0^{(\sigma_3)^2}\right) \left(G_2^2+G_3^2\right)
\nonumber\\&&\quad\quad\quad\quad
+\frac{g_3^2}{12}\, \left[ 
\left(A_0^{(\lambda_1)^2}+A_0^{(\lambda_2)^2}+A_0^{(\lambda_3)^2}\right) \left( F_1^2 +F_2^2 + G_1^2 + G_3^2 \right)
+\left(A_0^{(\lambda_4)^2}+A_0^{(\lambda_5)^2}\right) \left( F_1^2 +F_3^2 + G_1^2 + G_3^2 \right)
\right.\nonumber\\&&\quad\quad\quad\quad\quad\quad
+\left(A_0^{(\lambda_6)^2}+A_0^{(\lambda_7)^2}\right) \left( F_2^2 +F_3^2 \right)
+\frac{1}{3}A_0^{(\lambda_8)^2}\,\left( F_1^2 +F_2^2 +4\,F_3^2+ G_1^2 + G_3^2 \right)
\nonumber\\&&\quad\quad\quad\quad\quad\quad
\left.
+\frac{2}{\sqrt{3}} A_0^{(\lambda_3)^2}A_0^{(\lambda_8)^2}\left(F_1^2 - F_2^2 + G_1^2 + G_3^2\right)
\right]
\nonumber\\&&\quad\quad\quad\quad
-\frac{g_1\,g_2}{18} A_0^{(\sigma_3)}A_0^{(Y)} \left(3\,G_2^2+G_3^2\right)
-\frac{g_2\,g_3}{6\,\sqrt{3}} A_0^{(\sigma_3)} \left(\sqrt{3}\,A_0^{(\lambda_3)} + A_0^{(\lambda_8)}\right)G_3^2 
\nonumber\\&&\quad\quad\quad\quad
+ \frac{g_1\, g_3}{18\,\sqrt{3}} A_0^{(Y)} \left[ \sqrt{3} A_0^{(\lambda_3)} \left(4\,F_1^2 + 2\,F_2^2 - 2\,G_1^2 + G_3^2\right) + A_0^{(\lambda_8)} \left(4\,F_1^2 - 2\,F_2^2 + 4\,F_3^2 - 2\,G_1^2 + G_3^2\right) \right]
\nonumber\\&&\quad\quad\quad\quad
+\frac{1}{9}\left(4\,g_1\,A_0^{(Y)} +3\, g_3\,A_0^{(\lambda_3)} +\sqrt{3}\, g_3\,A_0^{(\lambda_8)}\right)F_1^2\,\Sigma_1'
-\frac{1}{9}\left(2\,g_1\,A_0^{(Y)} +3\, g_3\,A_0^{(\lambda_3)} -\sqrt{3}\, g_3\,A_0^{(\lambda_8)}\right)F_2^2\,\Sigma_2' 
\nonumber\\&&\quad\quad\quad\quad
-\frac{2}{9} \left(g_1\,A_0^{(Y)} + \sqrt{3}\, g_3\,A_0^{(\lambda_8)}\right) F_3^2\,\Sigma_3'
-\frac{1}{9}\left(2\,g_1\,A_0^{(Y)} -3\, g_3\,A_0^{(\lambda_3)} -\sqrt{3}\, g_3\,A_0^{(\lambda_8)}\right)G_1^2\,\tilde{\Sigma}_1'   
\nonumber\\&&\quad\quad\quad\quad+ \frac{1}{3} \left(g_1\,A_0^{(Y)} - g_2\,A_0^{(\sigma_3)}\right) G_2^2\,\tilde{\Sigma}_2'
-\frac{1}{9}\left(g_1\,A_0^{(Y)} -3\,g_2\, A_0^{(\sigma_3)} + 3\, g_3\,A_0^{(\lambda_3)} +\sqrt{3}\, g_3\,A_0^{(\lambda_8)}\right)G_3^2\,\tilde{\Sigma}_3'\Bigg\}\,,
\label{actbg2a}
\end{eqnarray}
where $A_\mu^{(G)}$ is the VEV of the gauge field corresponding to generator $G$. Integrating out the non-dynamical $A_0$ fields, the vector fields corresponding to non-diagonal operators are constrained to vanish. The ones of diagonal components are
\begin{eqnarray}
A_0^{(Y)} &=& \frac{1}{g_1} \left[ \frac{2\,F_2^2\,F_3^2 - F_1^2 \left(F_2^2+F_3^2\right)}{F_2^2 F_3^2+F_1^2 \left(F_2^2 +F_3^2\right)}\Sigma' - \frac{2\,G_2^2\,G_3^2 - G_1^2 \left(G_2^2+G_3^2\right)}{G_2^2 G_3^2+G_1^2 \left(G_2^2 +G_3^2\right)}\tilde{\Sigma}' - \Delta \Sigma_1'\right] \,,\nonumber\\
A_0^{(\sigma_3)} &=& \frac{1}{g_2} \left[ \frac{2\,F_2^2\,F_3^2 - F_1^2 \left(F_2^2+F_3^2\right)}{F_2^2 F_3^2+F_1^2 \left(F_2^2 +F_3^2\right)}\Sigma' + \frac{3\,G_1^2\,\left(G_2^2 - G_3^2\right)}{G_2^2 G_3^2+G_1^2 \left(G_2^2 +G_3^2\right)}\tilde{\Sigma}' - \Delta \Sigma_2'\right] \,,\nonumber\\
A_0^{(\lambda_3)} &=& \frac{1}{g_3} \left[ \frac{F_2^2\,F_3^2 + F_1^2 \left(F_2^2 - 2\,F_3^2\right)}{F_2^2 F_3^2+F_1^2 \left(F_2^2 +F_3^2\right)}\Sigma' + \frac{2\,G_2^2 G_3^2-G_1^2\left(G_2^2 + G_3^2\right)}{G_2^2 G_3^2+G_1^2 \left(G_2^2 +G_3^2\right)}\tilde{\Sigma}' - \Delta \Sigma_3'\right] \,,\nonumber\\
A_0^{(\lambda_8)} &=& \frac{1}{\sqrt{3}\,g_3} \left[ \frac{F_2^2\,F_3^2 - F_1^2 \left(5\,F_2^2- 4\,F_3^2 \right)}{F_2^2 F_3^2+F_1^2 \left(F_2^2 +F_3^2\right)}\Sigma' + \frac{2\,G_2^2 G_3^2-G_1^2\left(G_2^2 + G_3^2\right)}{G_2^2 G_3^2+G_1^2 \left(G_2^2 +G_3^2\right)}\tilde{\Sigma}' - \Delta \Sigma_4'\right] \,,
\end{eqnarray}
where,
\begin{eqnarray}
\Sigma = \frac{1}{3} \left(\Sigma_1 + \Sigma_2 + \Sigma_3\right) 
&\,,&
\tilde{\Sigma} = \frac{1}{3} \left(\tilde{\Sigma}_1 + \tilde{\Sigma}_2 + \tilde{\Sigma}_3\right) 
\,,\nonumber\\
\Delta\Sigma_1 = \frac{1}{3} \left(2\,\Sigma_1 - \Sigma_2 - \Sigma_3 - 2\,\tilde{\Sigma}_1 + \tilde{\Sigma}_2 + \tilde{\Sigma}_3\right) 
&\,,&
\Delta\Sigma_2 = \frac{1}{3} \left(2\,\Sigma_1 - \Sigma_2 - \Sigma_3 - 3\,\tilde{\Sigma}_2 +3\, \tilde{\Sigma}_3\right) 
\,,\nonumber\\
\Delta\Sigma_3 = \frac{1}{3} \left(\Sigma_1 - 2\,\Sigma_2 + \Sigma_3 + 2\,\tilde{\Sigma}_1 - \tilde{\Sigma}_2 - \tilde{\Sigma}_3\right) &\,,&
\Delta\Sigma_4 = \frac{1}{3} \left(\Sigma_1 + 4\, \Sigma_2 - 5\,\Sigma_3 + 2\,\tilde{\Sigma}_1 - \tilde{\Sigma}_2 - \tilde{\Sigma}_3\right) 
\,,
\end{eqnarray}

We replace the vector fields in (\ref{actbg2a}) using the above constraints, and obtain the following background equations:
\begin{eqnarray}
F_1''&+& \left(m_1^2\,R^2 - \frac{R''}{R}\right)F_1 + \frac{g_1^2}{54}\,F_1 \left(4\,F_1^2- 2\, F_2^2 -2\,F_3^2 - 2\,G_1^2 +3\,G_2^2 -G_3^2\right) \nonumber\\
& + & \frac{g_3^2}{36}\,F_1 \left(2\,F_1^2 - F_2^2 -F_3^2 + 2\,G_1^2 -2\,G_3^2\right) -\frac{9\,F_1\,F_2^4\,F_3^4}{\left[F_2^2 F_3^2+F_1^2 \left(F_2^2+F_3^2\right)\right]^2} \Sigma'^2 = 0 
\,,\nonumber\\
F_2''&+& \left(m_2^2\,R^2 - \frac{R''}{R}\right)F_2 - \frac{g_1^2}{108}\,F_2 \left(4\,F_1^2- 2\, F_2^2 -2\,F_3^2 - 2\,G_1^2 +3\,G_2^2 -G_3^2\right) \nonumber\\
& + & \frac{g_3^2}{36}\,F_2 \left(-F_1^2 +2\, F_2^2 -F_3^2 - G_1^2 + G_3^2\right) -\frac{9\,F_1^4\,F_2\,F_3^4}{\left[F_2^2 F_3^2+F_1^2 \left(F_2^2+F_3^2\right)\right]^2} \Sigma'^2 = 0 
\,,\nonumber\\
F_3''&+& \left(m_3^2\,R^2 - \frac{R''}{R}\right)F_3 - \frac{g_1^2}{108}\,F_3 \left(4\,F_1^2- 2\, F_2^2 -2\,F_3^2 - 2\,G_1^2 +3\,G_2^2 -G_3^2\right) \nonumber\\
& + & \frac{g_3^2}{36}\,F_3 \left(-F_1^2 - F_2^2 +2\,F_3^2 - G_1^2 + G_3^2\right) -\frac{9\,F_1^4\,F_2^4\,F_3}{\left[F_2^2 F_3^2+F_1^2 \left(F_2^2+F_3^2\right)\right]^2} \Sigma'^2 = 0 \,,\nonumber\\
G_1''&+& \left(m_4^2\,R^2 - \frac{R''}{R}\right)G_1 - \frac{g_1^2}{108}\,G_1 \left(4\,F_1^2- 2\, F_2^2 -2\,F_3^2 - 2\,G_1^2 +3\,G_2^2 -G_3^2\right) \nonumber\\
& + & \frac{g_3^2}{36}\,G_1 \left(2\,F_1^2 - F_2^2 - F_3^2 +2\, G_1^2 -2\, G_3^2\right) -\frac{9\,G_1\,G_2^4\,G_3^4}{\left[G_2^2 G_3^2+G_1^2 \left(G_2^2+G_3^2\right)\right]^2} \tilde{\Sigma}'^2 = 0 
\,,\nonumber\\
G_2''&+& \left(m_5^2\,R^2 - \frac{R''}{R}\right)G_2 + \frac{g_1^2}{72}\,G_2 \left(4\,F_1^2- 2\, F_2^2 -2\,F_3^2 - 2\,G_1^2 +3\,G_2^2 -G_3^2\right) \nonumber\\
& + & \frac{g_2^2}{24}\,G_2 \left(G_2^2 - G_3^2\right) -\frac{9\,G_1^4\,G_2\,G_3^4}{\left[G_2^2 G_3^2+G_1^2 \left(G_2^2+G_3^2\right)\right]^2} \tilde{\Sigma}'^2 = 0 
\,,\nonumber\\
G_3''&+& \left(m_6^2\,R^2 - \frac{R''}{R}\right)G_3 - \frac{g_1^2}{216}\,G_3 \left(4\,F_1^2- 2\, F_2^2 -2\,F_3^2 - 2\,G_1^2 +3\,G_2^2 -G_3^2\right) \nonumber\\
& - & \frac{g_2^2}{24}\,G_3 \left(G_2^2 - G_3^2\right) 
+\frac{g_3^2}{36}\,G_3 \left(-2\,F_1^2 + F_2^2 + F_3^2 -2\, G_1^2 + 2\, G_3^2\right) - \frac{9\,G_1^4\,G_2^4\,G_3}{\left[G_2^2 G_3^2+G_1^2 \left(G_2^2+G_3^2\right)\right]^2} \tilde{\Sigma}'^2 = 0 \,,\nonumber\\
&&\left(\frac{F_1^2\,F_2^2\,F_3^2}{F_2^2\,F_3^2+F_1^2\left(F_2^2+F_3^2\right)}\,\Sigma'\right)' = 0 \,,\quad\quad
\left(\frac{G_1^2\,G_2^2\,G_3^2}{G_2^2\,G_3^2+G_1^2\left(G_2^2+G_3^2\right)}\,\tilde{\Sigma}'\right)' = 0\,,
\label{eomf1f2f3}
\end{eqnarray}
where the equations of motion are independent of the gauge variant combinations $\Delta\Sigma_i$. Finally, with the definitions
\begin{eqnarray}
&&F_1 \equiv F + 2\,\Delta F_1 \,,\quad
F_2 \equiv F - \Delta F_1 +\Delta F_2 \,,\quad
F_3 \equiv F - \Delta F_1 -\Delta F_2 \,,\nonumber\\
&&G_1 \equiv G + 2\,\Delta G_1 \,,\quad
G_2 \equiv G - \Delta G_1 +\Delta G_2 \,,\quad
G_3 \equiv G - \Delta G_1 -\Delta G_2 \,,
\end{eqnarray}
and applying the expansion (\ref{2fdapp}), along with the assumptions $\Delta F'_i/\Delta F_i =\mathcal{O}(\epsilon)$, $\Delta G'_i/\Delta G_i =\mathcal{O}(\epsilon)$ (see the discussion in Section \ref{sec:1fdBG}), the equations of motion (\ref{eomf1f2f3}) give (\ref{ddexp2}) and (\ref{hexp2}).


\section{Quadratic action for the $u^c d^c d^c+Q L d^c$ example}\label{AppD}
In this appendix, we summarize the calculations of the quadratic action for the $u^c d^c d^c+Q L d^c$ perturbations of Section \ref{sec:2fdPER}. After fixing the unitary gauge, there are in total, 64 degrees of freedom, as well as 12 non-dynamical degrees. As mentioned in the main text, this system is very complicated, but it is possible to pick out decoupled subsystems. We study the spectra and adiabaticity conditions for each of these separately in the subsections below. In this following, we immediately integrate out the VEV of the non-dynamical degrees $\langle A_0^{(G)} \rangle$, so that $A_\mu^{(G)}$ denotes the perturbations to the gauge field corresponding to generator $G$.

\subsection{Subsystem $S_\perp^{(\rm non~diag)}$: Transverse vectors - non-diagonal generators }\label{appD1}

One of the immediate sub-systems of the complete quadratic action is the part containing the transverse vector degrees corresponding to non diagonal generators. These decouple from the rest, as well as from each other, each with a very similar action differing only in the mass terms:
\begin{equation}
S_\perp^{(G)} = \frac{1}{2} \int d^4 x \left[A_i^{(G)\prime}A_i^{(G)\prime}-\left(\partial_i\,A_j^{(G)}\right) \left(\partial_i\,A_j^{(G)}\right) - m^2_{(G)}A_i^{(G)} A_i^{(G)}\right]\,,
\label{transact}
\end{equation}
where all the masses are of order VEV,
\begin{eqnarray}
&&m^2_{(\lambda_1)} = m^2_{(\lambda_2)} = \frac{g_3^2}{12}\,\left(F_1^2+F_2^2+G_1^2+G_3^2\right) \,,\quad\quad
m^2_{(\lambda_4)} = m^2_{(\lambda_5)} = \frac{g_3^2}{12}\,\left(F_1^2+F_3^2+G_1^2+G_3^2\right) \,,\nonumber\\
&&m^2_{(\lambda_6)} = m^2_{(\lambda_7)} = \frac{g_3^2}{12}\,\left(F_2^2+F_3^2\right) \,,\quad\quad\quad\quad\quad\quad\quad
m^2_{(\sigma_1)} = m^2_{(\sigma_2)} = \frac{g_2^2}{12}\,\left(G_2^2+G_3^2\right) \,.
\end{eqnarray}

This part of the action decouples $16$ dynamical degrees of freedom from the rest.

\subsection{Subsystem $S_\perp^{(\rm diag)}$: Transverse vectors - diagonal generators}\label{appD2}

We now move on to the transverse part of the vectors corresponding to the diagonal generators. Although these are decoupled from the rest, they are coupled to each other. The action can be written in the form
\begin{equation}
S_\perp^{\rm (diag)} = \frac{1}{2} \int d^4x \left[ V^{T \prime}_i V^{\prime}_i - \left(\partial_i V^{T}_j\right)\left(\partial_i V_j\right) - V^T_i\,M^2_V V_i \right]\,,
\end{equation}
where
\begin{equation}
V_i \equiv  \left( A_i^{(Y)} \,,\; A_i^{(\sigma_3)} \,,\; A_i^{(\lambda_3)} \,,\; A_i^{(\lambda_8)} \right)\,.
\end{equation}
The mixing of the modes are due to the non-diagonal mass matrix components
\begin{equation}
\begin{array}{ll}
\left(M^2_V\right)_{11} = \frac{g_1^2}{108} \left(16\,F_1^2 + 4\, F_2^2 + 4\,F_3^2 +4\, G_1^2 +9\,G_2^2 + G_3^2 \right)
\,,&
\left(M^2_V\right)_{23} = -\frac{g_2 g_3}{12} G_3^2
\,,\\
\left(M^2_V\right)_{12} = -\frac{g_1 g_2}{36} \left(3\,G_2^2 +G_3^2\right)
\,,&
\left(M^2_V\right)_{24} = -\frac{g_2 g_3}{12\,\sqrt{3}} G_3^2
\,,\\
\left(M^2_V\right)_{13} = \frac{g_1 g_3}{36} \left(4\,F_1^2 + 2\,F_2^2 - 2\,G_1^2 + G_3^2 \right)
\,,&
\left(M^2_V\right)_{33} = \frac{g_3^2}{12} \left(F_1^2 + F_2^2 + G_1^2 + G_3^2\right)
\,,\\
\left(M^2_V\right)_{14} = \frac{g_1 g_3}{36\,\sqrt{3}} \left(4\,F_1^2 - 2\,F_2^2 + 4\, F_3^2 - 2\,G_1^2 + G_3^2 \right)
\,,&
\left(M^2_V\right)_{34} = \frac{g_3^2}{12\,\sqrt{3}} \left(F_1^2 - F_2^2 + G_1^2 + G_3^2\right)
\,,\\
\left(M^2_V\right)_{22} = \frac{g_2^2}{12} \left(G_2^2 +G_3^2\right)
\,,&
\left(M^2_V\right)_{44} = \frac{g_3^2}{36} \left(F_1^2 + F_2^2 + 4\,F_3^2 + G_1^2 + G_3^2\right)
\,,
\end{array}
\end{equation}
However, as discussed in Section \ref{sec:formalism}, this mixing does not result in a non-perturbative production of quanta. All the gauge symmetries are broken, so the eigenvalues of the mass matrix are all of order $1$ in an $\epsilon$ expansion. A straightforward way to verify this is the calculation of the determinant of $M^2_V$  at order $\epsilon^0$, which turns out to be non-zero. 

This part of the action decouples $8$ more dynamical degrees of freedom.

\subsection{Subsystems $S_{c_2}$ and $S_{c_3}$ : Scalar modes decoupled from the gauge fields}\label{appD3}
The part of the action containing perturbations of $c_2$ is
\begin{equation}
 S_{c_2} = \frac{1}{2}\int d^4 x \Bigg[ \delta_{17}^{\prime\,2} + \delta_{18}^{\prime\,2} - (\partial_i \delta_{17})(\partial_i \delta_{17})- (\partial_i \delta_{18})(\partial_i \delta_{18}) + m^2_{c_2} \left( \delta_{17}^2 + \delta_{18}^2\right)
+2\,K_{c_2}\left(\delta_{17}' \delta_{18} - \delta_{18}' \delta_{17}\right)\Bigg]\,,
\end{equation}
where 
\begin{eqnarray}
K_{c_2} &=& \frac{-2\,F_1^2 F_2^2 + (F_1^2 +F_2^2)F_3^2}{F_2^2 F_3^2 + F_1^2 (F_2^2+F_3^2)}\Sigma' +
\frac{G_1^2 G_2^2 - 2\,(G_1^2 +G_2^2)G_3^2}{G_2^2 G_3^2 + G_1^2 (G_2^2+G_3^2)}\tilde{\Sigma}'\,,\nonumber\\
m^2_{c_2} &=& m_6^2\,R^2 + \frac{y_d^2}{6}F_2^2 +\frac{g_1^2}{216} \left(-4\,F_1^2 +2\, F_2^2 +2\,F_3^2 +2\,G_1^2-3\,G_2^2 +G_3^2\right) \nonumber\\
&& +\frac{g_2^2}{24}\,\left(G_2^2 - G_3^2\right) +\frac{g_3^2}{36}\left(F_1^2 - 2\,F_2^2 + F_3^2 +G_1^2 - G_3^2 \right) - K_{c_2}^2\,.
\end{eqnarray}
With these, the frequency matrix $\Omega_4^2$ defined in Section \ref{sec:formalism} is diagonal and degenerate with eigenvalues $\omega^2 = k^2 +m_{c_2}^2+K_{c_2}^2$. For non-zero $y_d$, these modes are heavy, so the components of adiabaticity matrix $\mathcal{A}$ (eq. \ref{adia}) are at least of order $\epsilon$. Therefore, non-perturbative effects are suppressed for this system.

The second such system that does not couple to the gauge fields consists of the perturbations to $c_3$, with an identical action with $(\delta_{17},\delta_{18}) \rightarrow (\delta_{19} ,\delta_{20})\,,$ $y_d \rightarrow 0$ and $F_2\leftrightarrow F_3$. The main difference of this system from the previous one is the lack of $\mathcal{O}(\epsilon^0)$ terms in the frequency, that is, the eigenmodes of this system are light, with
\begin{equation}
\omega^2 = \left[k^2+ \left(\tilde{m}^2 + \delta m_2^2 -\frac{3\,\delta m_3^2}{2}+\frac{\delta m_4^2}{2}\right)R^2\right]\,,
\end{equation}
Again, due to the degeneracy of the states, the condition (\ref{adiabatic}) cannot be satisfied. Also, the eigenfrequencies evolve adiabatically, so we conclude that there is no non-perturbative production from these systems.

These two parts decouple four more dynamical degrees of freedom.

\subsection{Subsystem $S_{e,\,c_1}$: Perturbations to $e$ and $c_1$}\label{appD4}
This subsystem consists of perturbations to $e$ and $c_1$, coupled to the longitudinal vector degrees of non-diagonal generators of $SU(2)$. The action, in Fourier space, is 
\begin{eqnarray}
S_{e,\,c_1} &=& \frac{1}{2} \int d^3k \,d\eta \Bigg\{
\vert \delta_{15}' \vert^2 + \vert \delta_{16}' \vert^2 + k^2 \left( \vert A_0^{(\sigma_1)} - A_L^{(\sigma_1)\,\prime} \vert^2   + \vert A_0^{(\sigma_2)} - A_L^{(\sigma_2)\,\prime} \vert^2 \right)  \nonumber\\
&&\quad\quad\quad\quad\quad +\frac{g_2^2}{12}\left(G_2^2+G_3^2\right) \left[ \vert A_0^{(\sigma_1)} \vert^2 +\vert A_0^{(\sigma_2)} \vert^2 - k^2 \left(\vert  A_L^{(\sigma_1)} \vert^2 + \vert A_L^{(\sigma_2)} \vert^2\right) \right] \nonumber\\
&&\quad\quad\quad\quad\quad 
- \Bigg[ k^2 + \frac{\left( m_5^2 G_2^2 + m^2_6 G_3^2\right)R^2}{G_2^2+G_3^2} - \frac{\left(G_3 G_2' - G_2 G_3'\right)^2}{\left(G_2^2+G_3^2\right)^2}\nonumber\\
&&\quad\quad\quad\quad\quad\quad\quad  
+ \frac{g_1^2 }{216 } \frac{(3\,G_2^2 - G_3^2)\left(4\,F_1^2-2\,F_2^2 - 2\,F_3^2 -2\,G_1^2 +3\,G_2^2 -G_3^2 \right)}{( G_2^2+G_3^2)}  \nonumber\\
&&\quad\quad\quad\quad\quad\quad\quad 
+\frac{g_2^2}{24}\, \frac{G_2^4 + 6\,G_2^2 G_3^2 +G_3^4}{G_2^2+G_3^2} +\frac{g_3^2}{36}\, \frac{G_3^2 \left(-2\,F_1^2 + F_2^2 +F_3^2 -2\,G_1^2 +2\,G_3^2\right)}{G_2^2+G_3^2}  \nonumber\\
&&\quad\quad\quad\quad\quad\quad\quad
-\left( \frac{- 2\,F_2^2 F_3^2 + F_1^2 \left(F_2^2 +F_3^2\right)}{F_2^2 F_3^2 + F_1^2 \left(F_2^2 +F_3^2\right)} \Sigma'-\frac{3\,G_1^2 \left(G_2^2 - G_3^2 \right)}{G_2^2  G_3^2 +G_1^2 \left(G_2^2 + G_3^2\right)} \,\tilde{\Sigma}'\right)^2
\nonumber\\
&&\quad\quad\quad\quad\quad\quad\quad  
- \left(\frac{3\,G_1^2 G_2 G_3}{G_2^2  G_3^2 +G_1^2 \left(G_2^2 + G_3^2\right)}\tilde{\Sigma}'\right)^2
\Bigg]\left( |\delta_{15}|^2 +|\delta_{16}|^2 \right)  \nonumber\\
&&\quad\quad\quad\quad\quad\quad\quad
+\left(\delta_{16}^\star \delta_{15}'-\delta_{15}^\star \delta_{16}' + {\rm c.c.}\right)
\left( \frac{- 2\,F_2^2 F_3^2 + F_1^2 \left(F_2^2 +F_3^2\right)}{F_2^2 F_3^2 + F_1^2 \left(F_2^2 +F_3^2\right)} \Sigma'-\frac{3\,G_1^2 \left(G_2^2 - G_3^2 \right)}{G_2^2  G_3^2 +G_1^2 \left(G_2^2 + G_3^2\right)} \,\tilde{\Sigma}'\right)\nonumber\\
&&\quad\quad\quad\quad\quad 
+\frac{g_2}{2\,\sqrt{3}\,\sqrt{G_2^2+G_3^2}} \left(A_0^{(\sigma_1)}\delta_{16}^\star - A_0^{(\sigma_2)}\delta_{15}^\star +{\rm c.c}\right)\nonumber\\
&&\quad\quad\quad\quad\quad\quad\quad\times
\left( \frac{\left[- 2\,F_2^2 F_3^2 + F_1^2 \left(F_2^2 +F_3^2\right)\right]\left(G_2^2-G_3^2\right)}{F_2^2 F_3^2 + F_1^2 \left(F_2^2 +F_3^2\right)} \Sigma'-\frac{3\,G_1^2 \left(G_2^2 + G_3^2 \right)^2}{G_2^2  G_3^2 +G_1^2 \left(G_2^2 + G_3^2\right)} \,\tilde{\Sigma}'\right)\nonumber\\
&&\quad\quad\quad\quad\quad 
-\frac{g_2\,\left(G_2^2-G_3^2\right) \left(G_2 G_2'+G_3 G_3'\right)}{2\,\sqrt{3}\left(G_2^2+G_3^2\right)^{3/2}}  \left(A_0^{(\sigma_1)}\delta_{15}^\star + A_0^{(\sigma_2)}\delta_{16}^\star +{\rm c.c}\right) \nonumber
\end{eqnarray}
\begin{eqnarray}
&&\quad\quad\quad\quad\quad
+\frac{g_2\,\left(G_2^2-G_3^2\right)}{2\,\sqrt{3} \sqrt{G_2^2+G_3^2}}  
\left[\left( A_0^{(\sigma_1)}\delta_{15}^{\star\,\prime} + A_0^{(\sigma_2)}\delta_{16}^{\star\,\prime}\right) - k^2 \left( A_L^{(\sigma_1)}\delta_{15}^{\star} + A_L^{(\sigma_2)}\delta_{16}^{\star}\right) +{\rm c.c.}\right] \Bigg\}\,.
\end{eqnarray}
Next, we integrate out the non-dynamical degrees $A_0^{(\sigma_1)}$ and $A_0^{(\sigma_2)}$, then apply the prescription of Appendix \ref{AppA} by expanding in $\epsilon$ series, to finally get the form (\ref{canform2}). In the $\epsilon$ expansion, the mixing matrix $K_4$ is $\mathcal{O}(\epsilon^2)$, and the frequency matrix $\Omega_4^2$ is in block diagonal form, 
\begin{equation}
\Omega_4^2 = \left(\begin{array}{lr}
            \tilde{\Omega}^2 & \mathcal{O}(\epsilon^3) \\
            \mathcal{O}(\epsilon^3) & \tilde{\Omega}^2 
           \end{array}\right)
\,,
\label{omeform4}
\end{equation}
where $\tilde{\Omega}^2$ is a $2\times2$ matrix. In other words, this action with four degrees can be separated to two identical systems at this approximation order. The components of the frequency matrix are
\begin{eqnarray}
\tilde{\Omega}^2_{11} =\tilde{\Omega}^2_{22} &=& \frac{g_2^2\,G_0^2}{6} +\left\{ \frac{g_2^2 G_0 G_1}{3}\right\} +\left[ k^2 + \frac{1}{2}\,\tilde{m}^2 R^2+ \frac{3}{2}\,\tilde{\Sigma}_0^{\prime\,2}+ \frac{g_2^2 }{6}\left(G_1^2+2\,G_0G_2\right) +\frac{g_2^2R^2}{3}\left(-\frac{\delta m_1^2}{g_1^2}+\frac{2\,\delta m_3^2}{g_3^2}\right)\right]\,,\nonumber\\
\tilde{\Omega}^2_{12}  &=& \left[-\frac{1}{2}\tilde{m}^2\,R^2 - \frac{3}{2}\,\tilde{\Sigma}_0^{\prime\,2}\right]\,.
\end{eqnarray}
The eigenfrequencies of this system consists of a pair of
\begin{eqnarray}
\omega_A^2 &=& \frac{g_2^2\,G_0^2}{6} + \left\{\frac{g_2^2\,G_0G_1}{3} \right\} +
\left[k^2 + \frac{g_2^2}{6}\left(G_1^2+2\,G_0G_2\right) + \frac{g_2^2\,R^2}{3}\left(-\frac{\delta m_1^2}{g_1^2}+\frac{2\,\delta m_3^2}{g_3^2}\right)\right]\,,\nonumber\\
\omega_B^2 &=& \frac{g_2^2\,G_0^2}{6} + \left\{\frac{g_2^2\,G_0G_1}{3} \right\} +
\left[k^2 + \tilde{m}^2R^2+3\,\Sigma_0'^2+\frac{g_2^2}{6}\left(G_1^2+2\,G_0G_2\right) + \frac{g_2^2\,R^2}{3}\left(-\frac{\delta m_1^2}{g_1^2}+\frac{2\,\delta m_3^2}{g_3^2}\right)\right]\,.
\end{eqnarray}
The pairs of frequencies (A) correspond to the longitudinal vectors, as they coincide with the mass of $A_i^{(\sigma_1)}$ and $A_i^{(\sigma_2)}$ (\ref{transact}) at the given order. The remaining pair of modes (B) are the Higgses. As all four of the degrees are heavy, we conclude that this system does not contribute to non-perturbative decay.

This part decouples four more dynamical degrees from the rest.

\subsection{Subsystem $S_{b_{\bar{2}} ,\,s_{\bar{3}}}$: Perturbations to $b_{\bar{2}}$ and $s_{\bar{3}}$}\label{appD5}
This system is very similar to the previous one and consists of the perturbations to $b_{\bar{2}}$ and $s_{\bar{3}}$, along with the longitudinal components of vector fields corresponding to $SU(3)$ generators $\lambda_6$ and $\lambda_7$. The action in Fourier space is,
\begin{eqnarray}
S_{b_{\bar{2}} ,\,s_{\bar{3}}}&=& \frac{1}{2} \int d^3k \,d\eta \Bigg\{
\vert \delta_{13}' \vert^2 + \vert \delta_{14}' \vert^2 + k^2 \left( \vert A_0^{(\lambda_6)} - A_L^{(\lambda_6)\,\prime} \vert^2   + \vert A_0^{(\lambda_7)} - A_L^{(\lambda_7)\,\prime} \vert^2 \right)  \nonumber\\
&&\quad\quad\quad\quad\quad +\frac{g_3^2}{12}\left(F_2^2+F_3^2\right) \left[ \vert A_0^{(\lambda_6)} \vert^2 +\vert A_0^{(\lambda_7)} \vert^2 - k^2 \left(\vert  A_L^{(\lambda_6)} \vert^2 + \vert A_L^{(\lambda_7)} \vert^2\right) \right] \nonumber\\
&&\quad\quad\quad\quad\quad 
- \Bigg[ k^2 + \frac{\left( m_2^2 F_2^2 + m^2_3 F_3^2\right)R^2}{F_2^2+F_3^2} - \frac{\left(F_3 F_2' - F_2 F_3'\right)^2}{\left(F_2^2+F_3^2\right)^2}\nonumber\\
&&\quad\quad\quad\quad\quad\quad\quad  
- \frac{g_1^2 }{108} \left(4\,F_1^2-2\,F_2^2 - 2\,F_3^2 -2\,G_1^2 +3\,G_2^2 -G_3^2 \right)\nonumber\\
&&\quad\quad\quad\quad\quad\quad\quad 
-\frac{g_3^2}{36\,\left(F_2^2+F_3^2\right)}\,\left[\left(F_2^2 + F_3^2\right)\left(F_1^2+G_1^2-G_3^2\right) -2\,\left(	F_2^4+5\,F_2^2 F_3^2+F_3^4\right)\right]  
\nonumber\\
&&\quad\quad\quad\quad\quad\quad\quad  
- \frac{9\,F_1^4\left(F_2^4 - F_2^2 F_3^2 +F_3^4\right)}{\left[F_2^2  F_3^2 +F_1^2 \left(F_2^2 + F_3^2\right)\right]^2}\,\Sigma'^2
\Bigg]\left( |\delta_{13}|^2 +|\delta_{14}|^2 \right)\nonumber\\
&&\quad\quad\quad\quad\quad
+\left(\delta_{14}^\star \delta_{13}'-\delta_{13}^\star \delta_{14}' + {\rm c.c.}\right) 
\left( \frac{3\,F_1^2\left(F_2^2-F_3^2\right)}{F_2^2 F_3^2 + F_1^2 \left(F_2^2 +F_3^2\right)}\right) \Sigma'\nonumber\\
&&\quad\quad\quad\quad\quad
-\frac{g_3}{2\,\sqrt{3}}\left(A_0^{(\lambda_7)}\delta_{13}^\star - A_0^{(\lambda_6)}\delta_{14}^\star +{\rm c.c}\right) \left(\frac{3\,F_1^2\,\left(F_2^2+F_3^2\right)^{3/2}}{F_2^2 F_3^2 +F_1^2 \left(F_2^2+F_3^2\right)} \right) \Sigma'\nonumber
\end{eqnarray}
\begin{eqnarray}
&&\quad\quad\quad\quad\quad
-\frac{g_3\,\left(F_2^2-F_3^2\right) \left(F_2 F_2'+F_3 F_3'\right)}{2\,\sqrt{3}\left(F_2^2+F_3^2\right)^{3/2}}  \left(A_0^{(\lambda_6)}\delta_{13}^\star + A_0^{(\lambda_7)}\delta_{14}^\star +{\rm c.c}\right) \nonumber\\
&&\quad\quad\quad\quad\quad
+\frac{g_3\,\left(F_2^2-F_3^2\right)}{2\,\sqrt{3} \sqrt{F_2^2+F_3^2}}  
\left[\left( A_0^{(\lambda_6)}\delta_{13}^{\star\,\prime} + A_0^{(\lambda_7)}\delta_{14}^{\star\,\prime}\right) - k^2 \left( A_L^{(\lambda_6)}\delta_{13}^{\star} + A_L^{(\lambda_7)}\delta_{14}^{\star}\right) +{\rm c.c.}\right] \Bigg\}\,.
\end{eqnarray}
After integrating out the non-dynamical degrees $A_0^{(\lambda_6)}$ and $A_0^{(\lambda_7)}$, we repeat the steps in Appendix \ref{AppA} and obtain the action of form (\ref{canform}). As in the previous subsystem, the matrix $K_4$ is of order $\mathcal{O}(\epsilon^2)$ and $\Omega_4^2$ is of form (\ref{omeform4}), with 
\begin{eqnarray}
\tilde{\Omega}^2_{11} =\tilde{\Omega}^2_{22} &=& \frac{g_3^2\,F_0^2}{6} +\left\{ \frac{g_3^2 F_0 F_1}{3}\right\} +\left[ k^2 + \frac{1}{2}\,m^2 R^2+ \frac{3}{2}\,\Sigma_0^{\prime\,2}+ \frac{g_3^2 }{6}\left(F_1^2+2\,F_0F_2\right) +\frac{g_3^2R^2}{3}\left(\frac{\delta m_1^2}{g_1^2}+\frac{\delta m_2^2}{g_2^2}+\frac{\delta m_3^2}{g_3^2}\right)\right]\,,\nonumber\\
\tilde{\Omega}^2_{12}  &=& \left[-\frac{1}{2}m^2\,R^2 - \frac{3}{2}\,\Sigma_0^{\prime\,2}\right]\,.
\end{eqnarray}
The eigenfrequencies consist of two copies of
\begin{eqnarray}
\omega_A^2 &=& \frac{g_3^2\,F_0^2}{6} + \left\{\frac{g_3^2\,F_0F_1}{3} \right\} +
\left[k^2 + \frac{g_3^2}{6}\left(F_1^2+2\,F_0F_2\right) + \frac{g_3^2\,R^2}{3}\left(\frac{\delta m_1^2}{g_1^2}+ \frac{\delta m_2^2}{g_2^2}+ \frac{\delta m_3^2}{g_3^2}+\right)\right]\,,\nonumber\\
\omega_B^2 &=& \frac{g_3^2\,F_0^2}{6} + \left\{\frac{g_3^2\,F_0F_1}{3} \right\} +
\left[k^2 + m^2R^2+3\,\Sigma_0'^2+ \frac{g_3^2}{6}\left(F_1^2+2\,F_0F_2\right) + \frac{g_3^2\,R^2}{3}\left(\frac{\delta m_1^2}{g_1^2}+ \frac{\delta m_2^2}{g_2^2}+ \frac{\delta m_3^2}{g_3^2}+\right)\right]\,.
\end{eqnarray}
Again, $\omega_A^2$ coincides with the $\epsilon$ expansion of the frequency of $A_i^{(\lambda_6)}$ and $A_i^{(\lambda_7)}$ (\ref{transact}), so it corresponds to the two longitudinal components, whereas $\omega_B^2$ are the frequencies of the two Higgses. Neither of these heavy modes will contribute to non-perturbative production. 

This system decouples another four dynamical degrees from the rest.

\subsection{Subsystem $S_{\delta f,\,\delta g}$: Perturbations to VEVs and longitudinal vectors - diagonal generators}\label{appD6}

This system consists of the perturbations to the field components with non-zero VEVs, coupled to the longitudinal vectors corresponding to the diagonal generators, namely, $\delta f_i$ (6 degrees), $\delta g_i$ (2 degrees) and $A_L^{(Y)}$, $A_L^{(\sigma_3)}$, $A_L^{(\lambda_3)}$, $A_L^{(\lambda_8)}$.  The initial action is very long for presentation. Furthermore, even the zero order terms in $\epsilon$ approximation are too bulky, so we will describe the spectrum and comment on how we identify the flat direction excitations.

After integrating out the four non dynamical degrees, we need to expand the matrices in (\ref{initform}) as a series in $\epsilon$, and repeat the steps in Appendix \ref{AppA} to get the form (\ref{canform2}). In the end, we find that at the relevant expansion order, two systems of each $4$ degrees can be decoupled from the rest. Similar to the check we performed in Appendix \ref{appD2}, we calculate the determinants of their $\Omega_4^2+K^T_4 K_4$ matrix at order $\epsilon^0$, which shows that all eight of these modes are heavy, i.e. the system contains four Higgses and four longitudinal vector modes corresponding to the diagonal generators. 

What remain are two pairs of light modes, with leading order eigenfrequencies
\begin{equation}
\omega^2 = \left[k^2+m^2 R^2 - \frac{R''}{R} \right]\,, \quad
\tilde{\omega}^2 = \left[k^2+\tilde{m}^2 R^2 - \frac{R''}{R} \right]\,.
\end{equation}
This system is actually the analogue of the coupled system for the $2$ field toy model in Section \ref{AppB}. As in that case, the perturbations to the field components with VEVs are coupled to the longitudinal gauge fields of diagonal generators. The only light modes in this action are then clearly the perturbations of the two flat directions. Indeed, in the degenerate limit $\delta m_i = 0$, $F_0=F$, $G_0=G$, $\Sigma_0=\Sigma$, $\tilde{\Sigma}_0=\tilde{\Sigma}$, the action containing the combinations
\begin{eqnarray}
r \equiv \frac{F}{2} \left( \cos \Sigma \,\delta f_1 - \sin \Sigma \, \delta g_1\right) &\,,&\;
\sigma \equiv \frac{F}{2} \left( \sin \Sigma \,\delta f_1 + \cos \Sigma \, \delta g_1\right) \,,\nonumber\\
\tilde{r} \equiv \frac{G}{2} \left( \cos \tilde{\Sigma} \,\delta f_4 - \sin \tilde{\Sigma} \, \delta g_2\right) &\,,&\;
\tilde{\sigma} \equiv \frac{G}{2} \left( \sin \tilde{\Sigma} \,\delta f_4 + \cos \tilde{\Sigma} \, \delta g_2\right) \,,
\end{eqnarray}
immediately decouples from the rest and one recovers the action described above at the leading order in $\epsilon$ expansion.

To summarize, this subsystem consists of $8$ heavy fields which decouple from the rest, and do not give rise to non-perturbative decay as described in Section \ref{sec:formalism}. The remaining light modes correspond to flat direction excitations, with adiabatically evolving frequencies. Although there is a non-zero $\Gamma$ matrix, the light modes that mix are degenerate, so these do not contribute to production either.

This part of the action eliminates $12$ more of the dynamical degrees of freedom.

\subsection{Subsystem $S_{s_{\bar{1}} ,\,s_2,\,d_{\bar{2}}} $: Perturbations to $s_{\bar{1}}$, $s_2$ and $d_{\bar{2}}$} \label{appD7}
This subsystem consists of the perturbations to the field components $s_{\bar{1}}$, $s_2$ and $d_{\bar{2}}$, that is, $\delta_i$, $i\in[1,6]$, along with the longitudinal components of the vector fields corresponding to $SU(3)$ generators $\lambda_1$ and $\lambda_2$.  The decoupled action is
\begin{eqnarray}
S_{s_{\bar{1}} ,\,s_2,\,d_{\bar{2}}} &=& \frac{1}{2} \int d^4x \Bigg\{
\sum_{a=1}^6
\left[\delta_{a}^{\prime\, 2} - (\partial_i\delta_a)^2\right]+ (\partial_i A_L^{(\lambda_1)\,\prime})^2 + (\partial_i A_L^{(\lambda_2)\,\prime})^2 + (\partial_i A_0^{(\lambda_1)})^2 + (\partial_i A_0^{(\lambda_2)})^2
\nonumber\\&&\quad\quad\quad\quad\quad 
-2\, (\partial_i A_L^{(\lambda_1)\,\prime})(\partial_i A_0^{(\lambda_1)}) - 2\, (\partial_i A_L^{(\lambda_2)\,\prime})(\partial_i A_0^{(\lambda_2)})
\nonumber\\&&\quad\quad\quad\quad\quad 
-m_v^2 \left[ (\partial_i A_L^{(\lambda_1)})^2 + (\partial_i A_L^{(\lambda_2)})^2 -(A_0^{(\lambda_1)})^2 -(A_0^{(\lambda_2)})^2\right]
\nonumber\\&&\quad\quad\quad\quad\quad 
-\frac{g_3\,F_2'}{\sqrt{3}} \left( A_0^{(\lambda_1)} \delta_1 + A_0^{(\lambda_2)} \delta_4\right)
+\frac{g_3\,G_3'}{\sqrt{3}} \left( A_0^{(\lambda_1)} \delta_2 + A_0^{(\lambda_2)} \delta_5\right)
-\frac{g_3\,G_1'}{\sqrt{3}} \left( A_0^{(\lambda_1)} \delta_3 + A_0^{(\lambda_2)} \delta_6\right)
\nonumber\\&&\quad\quad\quad\quad\quad 
-m_{v \,s_{\bar{1}}}^2 \left( A_0^{(\lambda_1)} \delta_4 - A_0^{(\lambda_2)} \delta_1\right)
-m_{v \,s_2}^2 \left( A_0^{(\lambda_1)} \delta_5 - A_0^{(\lambda_2)} \delta_2\right)
-m_{v\,d_{\bar{2}}}^2 \left( A_0^{(\lambda_1)} \delta_6 - A_0^{(\lambda_2)} \delta_3\right)
\nonumber\\&&\quad\quad\quad\quad\quad 
+\frac{g_3\,F_2}{\sqrt{3}}\,\left( A_0^{(\lambda_1)} \delta_1'+ A_0^{(\lambda_2)} \delta_4' - (\partial_i A_L^{(\lambda_1)})(\partial_i \delta_1) - (\partial_i A_L^{(\lambda_2)})(\partial_i \delta_4)\right) -m_{s_{\bar{1}}}^2 \left(\delta_1^2 + \delta_4^2\right)
\nonumber\\&&\quad\quad\quad\quad\quad 
-\frac{g_3\,G_3}{\sqrt{3}}\,\left( A_0^{(\lambda_1)} \delta_2'+ A_0^{(\lambda_2)} \delta_5' - (\partial_i A_L^{(\lambda_1)})(\partial_i \delta_2) - (\partial_i A_L^{(\lambda_2)})(\partial_i \delta_5)\right) - m_{s_2}^2 \left(\delta_2^2 + \delta_5^2\right) 
\nonumber\\&&\quad\quad\quad\quad\quad 
+\frac{g_3\,G_1}{\sqrt{3}}\,\left( A_0^{(\lambda_1)} \delta_3'+ A_0^{(\lambda_2)} \delta_6' - (\partial_i A_L^{(\lambda_1)})(\partial_i \delta_3) - (\partial_i A_L^{(\lambda_2)})(\partial_i \delta_6)\right) - m_{d_{\bar{2}}}^2 \left(\delta_3^2 + \delta_6^2\right)
\nonumber\\&&\quad\quad\quad\quad\quad
+ \frac{(g_3^2-2\,y_d^2)}{6} F_2 G_3 \left(\delta_1 \delta_2 +\delta_4 \delta_5\right) 
+\frac{g_3^2}{6} F_2 G_1 \left(\delta_1 \delta_3 +\delta_4 \delta_6\right)
-\frac{g_3^2}{6} G_1 G_3 \left(\delta_2 \delta_3 +\delta_5 \delta_6\right)
\nonumber\\&&\quad\quad\quad\quad\quad
+2\,\chi_{s_{\bar{1}}} \left(\delta_1 \delta_4'-\delta_1'\delta_4\right) 
+2\,\chi_{s_2} \left(\delta_2 \delta_5'-\delta_2'\delta_5\right) 
+2\,\chi_{d_{\bar{2}}} \left(\delta_3 \delta_6'-\delta_3' \delta_6\right) \Bigg\}\,,
\end{eqnarray}
with the mass parameters,
\begin{eqnarray}
m_v^2 &\equiv& \frac{g_3^2}{12}\left(F_1^2+F_2^2+G_1^2+G_3^2\right)\,,\nonumber\\
m_{s_{\bar{1}}}^2 &\equiv& m_2^2 R^2 +\frac{y_d^2}{6}\,G_3^2 + \frac{g_1^2}{108}\,\left(-4\,F_1^2 +2\,F_2^2 +2\,F_3^2 +2\,G_1^2 -3\,G_2^2 +G_3^2\right)
\nonumber\\&&
+\frac{g_3^2}{36}\,\left(2\,F_1^2 +2\,F_2^2-F_3^2 +2\,G_1^2 -2\,G_3^2\right) - \chi_{s_{\bar{1}}}^2 - \frac{R''}{R}\,,\nonumber\\
m_{s_2}^2 &\equiv& m_6^2 R^2 +\frac{y_d^2}{6}\,F_2^2 + \frac{g_1^2}{216}\,\left(-4\,F_1^2 +2\,F_2^2 +2\,F_3^2 +2\,G_1^2 -3\,G_2^2 +G_3^2\right)
\nonumber\\&&
+\frac{g_2^2}{24}\,\left(-G_2^2 + G_3^2\right)+\frac{g_3^2}{36}\,\left(F_1^2 -2\,F_2^2+F_3^2 + G_1^2 + 2\,G_3^2\right) - \chi_{s_2}^2 - \frac{R''}{R}\,,\nonumber\\
m_{d_{\bar{2}}}^2 &\equiv& m_4^2 R^2  + \frac{g_1^2}{108}\,\left(-4\,F_1^2 +2\,F_2^2 +2\,F_3^2 +2\,G_1^2 -3\,G_2^2 +G_3^2\right)
\nonumber
\end{eqnarray}
\begin{eqnarray}
&&+\frac{g_3^2}{36}\,\left(-F_1^2 +2\,F_2^2-F_3^2 +2\,G_1^2 + G_3^2\right) - \chi_{d_{\bar{2}}}^2 - \frac{R''}{R}\,,
\end{eqnarray}
and the couplings, 
\begin{eqnarray}
\chi_{s_{\bar{1}}} &\equiv& \Sigma' - \frac{G_1^2 G_2^2 + G_1^2 G_3^2 -2\,G_2^2 G_3^2}{G_1^2 G_2^2 +G_1^2 G_3^2 +G_2^2 G_3^2}\,\tilde{\Sigma}' \,,\nonumber\\
\chi_{s_2} &\equiv& \frac{F_1^2 F_2^2 -2\,F_1^2 F_3^2 +F_2^2 F_3^2}{F_1^2 F_2^2 +F_1^2 F_3^2 +F_2^2 F_3^2}\,\Sigma' + \frac{2\,G_1^2 G_2^2 - G_1^2 G_3^2 +2\,G_2^2 G_3^2}{G_1^2 G_2^2 +G_1^2 G_3^2 +G_2^2 G_3^2}\,\tilde{\Sigma}' \,,\nonumber\\
\chi_{d_{\bar{2}}} &\equiv& \frac{F_1^2 F_2^2 -2\,F_1^2 F_3^2 +F_2^2 F_3^2}{F_1^2 F_2^2 +F_1^2 F_3^2 +F_2^2 F_3^2}\,\Sigma' - \tilde{\Sigma}' \,,\nonumber\\
m_{v\,s_{\bar{1}}}^2 &\equiv& \frac{g_3\,F_2}{\sqrt{3}} \left( \frac{F_1^2 F_2^2 +4\,F_1^2 F_3^2 +F_2^2 F_3^2}{F_1^2 F_2^2 +F_1^2 F_3^2 +F_2^2 F_3^2}\,\Sigma' - \frac{G_1^2 G_2^2 + G_1^2 G_3^2 - 2\,G_2^2 G_3^2}{G_1^2 G_2^2 +G_1^2 G_3^2 +G_2^2 G_3^2}\,\tilde{\Sigma}' \right)\,,\nonumber\\
m_{v\,s_2}^2 &\equiv& -\frac{g_3\,G_3}{\sqrt{3}} \left( \frac{F_1^2 F_2^2 -2\,F_1^2 F_3^2 +F_2^2 F_3^2}{F_1^2 F_2^2 +F_1^2 F_3^2 +F_2^2 F_3^2}\,\Sigma' + \frac{5\,G_1^2 G_2^2 - G_1^2 G_3^2 + 2\,G_2^2 G_3^2}{G_1^2 G_2^2 +G_1^2 G_3^2 +G_2^2 G_3^2}\,\tilde{\Sigma}' \right)\,,\nonumber\\
m_{v\,d_{\bar{2}}}^2 &\equiv& \frac{g_3\,G_1}{\sqrt{3}} \left( \frac{F_1^2 F_2^2 -2\,F_1^2 F_3^2 +F_2^2 F_3^2}{F_1^2 F_2^2 +F_1^2 F_3^2 +F_2^2 F_3^2}\,\Sigma' - \frac{G_1^2 G_2^2 + G_1^2 G_3^2 +4 \,G_2^2 G_3^2}{G_1^2 G_2^2 +G_1^2 G_3^2 +G_2^2 G_3^2}\,\tilde{\Sigma}' \right)\,.
\end{eqnarray}

We then expand the fields in terms of plane waves and integrate out the non-dynamical degrees $A_0^{(\lambda_1)}$ and $A_0^{(\lambda_2)}$, recovering the form (\ref{initform}). As in the previous case, we apply the prescription in Appendix \ref{AppA} using $\epsilon$ expansion from the start. The resulting $\Omega^2_4+K_4^T K_4$ matrix is of the form
\begin{equation}
\Omega^2_4+K_4^T K_4= \left( 
\begin{array}{ll}
A & B\\
-B & A
\end{array}\right)\,,
\label{omform6}
\end{equation}
where $A$ and $B$ are $4\times4$ real matrices with symmetries $A^T = A$ and $B^T = -B$. The eigenvalues of such a matrix consist of two copies of the eigenvalues of the Hermitian matrix $A- i\,B$. A detailed discussion on diagonalization of this matrix form is given in Appendix \ref{AppE}. The eigenstates of this coupled system contains a pair of light modes,
\begin{eqnarray}
\omega_A^2 &=& \Bigg[ k^2 +\frac{3\,\left(G_0 F_0' -F_0 G_0'\right)^2}{2\,\left(F_0^2+G_0^2\right)^2} + \frac{3\,F_0^2 G_0^2 \left(\Sigma_0' -\tilde{\Sigma}_0'\right)^2}{2\,\left(F_0^2+G_0^2\right)^2 } +\frac{\left(\tilde{m}^2 F_0^2+m^2 G_0^2\right)R^2}{F_0^2+G_0^2} +\frac{\left(3\,\delta m_3^2+\delta m_4^2\right)\,R^2}{2}\Bigg]\,,
\end{eqnarray}
a pair of heavy modes with order $\epsilon^0$ term proportional to the Yukawa coupling,
\begin{eqnarray}
\omega_B^2 &=& \frac{y_d^2}{6} \left(F_0^2+G_0^2\right) + \left\{ \frac{y_d^2}{3} \left(F_0 F_1 +G_0 G_1 \right)\right\}  \nonumber\\
&&+ \Bigg[  k^2 +\frac{3\,\left(G_0 F_0' -F_0 G_0'\right)^2}{2\,\left(F_0^2+G_0^2\right)^2} + \frac{3\,F_0^2 G_0^2 \left(\Sigma_0' -\tilde{\Sigma}_0'\right)^2}{2\,\left(F_0^2+G_0^2\right)^2 }  +\frac{\left(\tilde{m}^2 F_0^2+m^2 G_0^2\right)R^2}{F_0^2+G_0^2}\nonumber\\
&&\quad\quad+\frac{y_d^2}{6}\left(F_1^2+G_1^2+2\,F_0 F_2 +2\,G_0 G_2 \right)
- y_d^2 R^2 \left(\frac{3\,\delta m_3^2+\delta m_4^2}{2\,y_d^2}+\frac{2\,\delta m_2^2}{3\,g_2^2}-\frac{\delta m_3^2+\delta m_4^2}{g_3^2}\right)\Bigg]\,,
\end{eqnarray}
a pair of Higgses,
\begin{eqnarray}
\omega_C^2 &=& \frac{g_3^2}{6} \left(F_0^2+G_0^2\right) + \left\{ \frac{g_3^2}{3} \left(F_0 F_1 +G_0 G_1 \right)\right\}  \nonumber\\
&&+ \Bigg[  k^2  + \frac{3\,\left(F_0^2 \Sigma_0'+ G_0^2 \tilde{\Sigma}_0'\right)^2}{\left(F_0^2+G_0^2\right)^2 }  +\frac{\left(m^2 F_0^2+\tilde{m}^2 G_0^2\right)R^2}{F_0^2+G_0^2}
\nonumber\\
&&\quad\quad+\frac{g_3^2}{6}\left(F_1^2+G_1^2+2\,F_0 F_2 +2\,G_0 G_2 \right)-\frac{g_3^2R^2}{6}\left(\frac{4\,\delta m_2^2}{g_2^2}+\frac{3\,\left(\delta m_3^2-\delta m_4^2\right)}{g_3^2}\right)
 \Bigg]\,,\nonumber\\
\end{eqnarray}
and finally, a pair of longitudinal vector components,
\begin{eqnarray}
\omega_D^2 &=& \frac{g_3^2}{6} \left(F_0^2+G_0^2\right) + \left\{ \frac{g_3^2}{3} \left(F_0 F_1 +G_0 G_1 \right)\right\}  \nonumber\\
&&+ \left[  k^2 +\frac{g_3^2}{6}\left(F_1^2+G_1^2+2\,F_0 F_2 +2\,G_0 G_2 \right)-\frac{g_3^2R^2}{6}\left(\frac{4\,\delta m_2^2}{g_2^2}+\frac{3\,\left(\delta m_3^2-\delta m_4^2\right)}{g_3^2}\right)\right]\,,\nonumber\\
\end{eqnarray}
coinciding with the masses of transverse vectors $A_i^{(\lambda_1)}$ and $A_i^{(\lambda_2)}$ from (\ref{transact}) at the given order in the expansion.
The two eigenvectors corresponding to each distinct eigenvalue at order $\epsilon^0$ are of the form,
\begin{equation}
V_{j,1} = \left( \begin{array}{r}   \cos\alpha_j \,{\bf w}_{j} \\ \sin\alpha_j \,{\bf w}_{j} \end{array} \right) \,,\quad\quad
V_{j,2} = \left( \begin{array}{r}   -\sin\alpha_j \,{\bf w}_{j} \\ \cos\alpha_j \,{\bf w}_{j} \end{array} \right) \,,\quad\quad
 (j = A,B,C,D)
\label{evecsyd}
\end{equation}
where $\alpha_j$ parameters are arbitrary rotations between the eigenvectors of a degenerate eigenvalue pair. Although there is an additional degeneracy at zero order between heavy modes ($C$) and ($D$), the bottom blocks of the eigenvalues turn out to be proportional to the upper blocks. The vectors ${\bf w}_j$ are, at order $\epsilon^0$,
\begin{eqnarray}
{\bf w}_{A} &=& \frac{F_0}{\sqrt{2\,(F_0^2+G_0^2)}}\left( 
\begin{array}{llll}
 0 & ,\, 1 & ,\, f_A (F_0,G_0) & ,\, f_A (i\,F_0 ,i \,G_0)
\end{array}
\right) \,,\nonumber\\
{\bf w}_{B} &=& \frac{F_0}{\sqrt{2\,(F_0^2+G_0^2)}}\left( 
\begin{array}{llll}
 0 & ,\, 1 & ,\, f_B (F_0,G_0) & ,\, f_B (i\,F_0 ,i \,G_0)
\end{array}
\right) \,,\nonumber\\
{\bf w}_{C} &=& \frac{G_0}{\sqrt{F_0^2+G_0^2}}\left( 
\begin{array}{llll}
 0 & ,\, 1 & ,\, f_C (F_0,G_0) & ,\, f_C (i\,F_0 ,i \,G_0)
\end{array}
\right) \,,\nonumber\\
{\bf w}_{D} &=& \left( 
\begin{array}{llll}
 1 & ,\, 0 & ,\, 0& ,\, 0
\end{array}
\right) \,,
\label{evecA}
\end{eqnarray}
where the functions are defined as
\begin{eqnarray}
f_A (F_0,G_0) &\equiv& -\frac{F_0^2+\sqrt{F_0^4 + 8\,G_0^4}}{2\,\sqrt{2}\,F_0 G_0}\,
\left( \frac{-F_0^2+2\,G_0^2+\sqrt{F_0^4 + 8\,G_0^4}}{\sqrt{F_0^4 + 8\,G_0^4}}\right)^{1/2}\,, \nonumber\\
f_B (F_0,G_0) &\equiv& \frac{G_0\left(3\,F_0^2-\sqrt{F_0^4 + 8\,G_0^4}\right)}{\sqrt{2}\,F_0 \left[
\left(-F_0^2+2\,G_0^2+\sqrt{F_0^4 + 8\,G_0^4}\right)\sqrt{F_0^4 + 8\,G_0^4}\right]^{1/2}
}\,\,, \nonumber\\
f_C (F_0,G_0) &\equiv& \frac{F_0}{\sqrt{2}\,G_0}\,
\left( \frac{-F_0^2+2\,G_0^2+\sqrt{F_0^4 + 8\,G_0^4}}{\sqrt{F_0^4 + 8\,G_0^4}}\right)^{1/2}\,.
\end{eqnarray}
Using the $K_4$ matrix at order $\epsilon$ and the above eigenvectors, we calculate $\mathcal{O}(\epsilon)$ terms of $\Gamma$ matrix. The non-zero components are
\begin{eqnarray}
\Gamma_{12} &=& -\alpha_A'+\frac{G_0^2  \, \Sigma_0' +F_0^2 \,\tilde{\Sigma}_0'}{F_0^2+G_0^2}  \,,\quad
\Gamma_{34} = -\alpha_B'-\frac{G_0^2  \, \Sigma_0' +F_0^2 \,\tilde{\Sigma}_0'}{F_0^2+G_0^2}  \,,\quad
\Gamma_{56} = -\alpha_C' \,,\quad \Gamma_{78}=-\alpha_D'\,,\nonumber\\
\Gamma_{15} &=& \Gamma_{26} = -\frac{F_0\,G_0}{\sqrt{2}(F_0^2+G_0^2)} \left[ \cos(\alpha_A-\alpha_C) \left(\frac{F_0'}{F_0} - \frac{G_0'}{G_0}\right) - \sin(\alpha_A - \alpha_C) \left(\Sigma_0'-\tilde{\Sigma}_0'\right)\right]\,,\nonumber\\
\Gamma_{16} &=& -\Gamma_{25} = -\frac{F_0\,G_0}{\sqrt{2}(F_0^2+G_0^2)} \left[ \sin(\alpha_A-\alpha_C) \left(\frac{F_0'}{F_0} - \frac{G_0'}{G_0}\right) + \cos(\alpha_A - \alpha_C) \left(\Sigma_0'-\tilde{\Sigma}_0'\right)\right]\,,\nonumber\\
\Gamma_{35} &=& \Gamma_{46} = -\frac{F_0\,G_0}{\sqrt{2}(F_0^2+G_0^2)} \left[ \cos(\alpha_B-\alpha_C) \left(\frac{F_0'}{F_0} - \frac{G_0'}{G_0}\right) + \sin(\alpha_B - \alpha_C) \left(\Sigma_0'-\tilde{\Sigma}_0'\right)\right]\,,\nonumber\\
\Gamma_{36} &=& -\Gamma_{45} = -\frac{F_0\,G_0}{\sqrt{2}(F_0^2+G_0^2)} \left[ \sin(\alpha_B-\alpha_C) \left(\frac{F_0'}{F_0} - \frac{G_0'}{G_0}\right) - \cos(\alpha_B - \alpha_C) \left(\Sigma_0'-\tilde{\Sigma}_0'\right)\right]\,,\nonumber\\
\label{gammayd}
\end{eqnarray}
We immediately see that the longitudinal vector modes $(7,8)$ are decoupled from the rest of the system and have only a mixing term between themselves, which does not contribute to production due their $\mathcal{O}(VEV)$ mass and degeneracy. Since the physical quantities are unaffected by the choice of $\alpha_i$ (see Appendix \ref{AppE} ), it is useful to study the $\Gamma$ matrix for a suitable choice.  As our main focus is to check for non-perturbative production, we look for a choice of these parameters which removes the non-adiabatic mixing of the eigenstates completely, that is, that makes the adiabaticity matrix $\mathcal{A}$ (eq \ref{adia}) to be of order $\epsilon$ at least. From the discussion in Section \ref{sec:formalism}, the only non-diagonal components of the adiabaticity matrix which has non-zero $\mathcal{O}(\epsilon^0)$ terms will be the ones which involve the mixing of the light mode $(1,2)$ to the Higgs $(5,6)$. If, in this system, the non-adiabatic rotation of the eigenstates is a spurious effect, we should be able to remove the $\Gamma_{15}$ and $\Gamma_{16}$ (or, equivalently, $\Gamma_{25}$ and $\Gamma_{26}$) components by proper choice of rotation parameters. However, from (\ref{gammayd}), we see that it is not possible to make both of these components zero simultaneously. Therefore, we conclude that the non-adiabatic mixing of the light modes $(1,2)$ to the heavy modes $(5,6)$ is a physical effect, which cannot be removed by exploiting the freedom in the eigenvectors. In addition, the light modes' eigenfrequency evolves non-adiabatically, so the diagonal condition (\ref{adiadiag}) may also contribute to production. 

This subsystem decouples a further 8 degrees of freedom.

\subsection{Subsystem $S_{b_{\bar{1}} ,\,s_3,\,d_{\bar{3}}}$: Perturbations to $s_{\bar{1}}$, $s_2$ and $d_{\bar{2}}$}\label{appD8}

This subsystem consists of the perturbations $\delta_i$, $i\in[7,12]$, along with the longitudinal components of the vector fields corresponding to $SU(3)$ generators $\lambda_4$ and $\lambda_5$. It is very similar to the one discussed in the previous subsection. In fact, the action for system $S_{b_{\bar{1}} ,\,s_3,\,d_{\bar{3}}}$ can be obtained by 
\begin{equation}
S_{b_{\bar{1}} ,\,s_3,\,d_{\bar{3}}}= S_{s_{\bar{1}} ,\,s_2,\,d_{\bar{2}}}\left(\delta_i \rightarrow \delta_{i+6} ,\, A_\mu^{(\lambda_i)} \rightarrow A_\mu^{(\lambda_{i+3})} ,\, F_2 \leftrightarrow F_3 ,\, m_2 \rightarrow m_3 ,\, y_d \rightarrow 0\right)\,.
\end{equation}
The calculations for the action proceed the same way as the previous case, up to the point where we have the action of the form (\ref{canform2}). When $\epsilon$ expansion is applied, $\Omega^2$ matrices can be related by
\begin{equation}
\Omega^2 _{b_{\bar{1}} ,\,s_3,\,d_{\bar{3}}}= \Omega^2_{s_{\bar{1}} ,\,s_2,\,d_{\bar{2}}}  \left(\delta m_4^2 \rightarrow -\delta m_4^2 ,\, y_d \rightarrow 0\right)\,,
\end{equation}
whereas the $K$ matrix, which is independent of $\delta m_4$ and $y_d$ stays the same.

Although we have the simple relation between this system and the previous one, the solution of the eigenvalue problem cannot be recovered by use of this correspondence. Specifically, the limit $y_d \rightarrow 0$ changes the picture dramatically in the light mode sector: a pair of the heavy modes from the previous case becomes light and along with the already existing pair of light modes, forms a fourfold degeneracy at zero order. On the other hand, the remaining pairs of two heavy modes do not undergo a modification and the eigenvectors and eigenvalues stay the same as before, with the exception of $\delta m_4^2 \rightarrow -\delta m_4^2$.

To summarize, the eigenfrequencies of the pairs of two light modes are
\begin{eqnarray}
\omega_A^2 &=& \Bigg[ k^2 +\left(\frac{3\,\left(G_0 F_0' -F_0 G_0'\right)^2}{2\,\left(F_0^2+G_0^2\right)^2} + \frac{3\,F_0^2 G_0^2 \left(\Sigma_0' -\tilde{\Sigma}_0'\right)^2}{2\,\left(F_0^2+G_0^2\right)^2 }\right)\left(1+\sqrt{1+\Delta M^2}\right)+\frac{\left(\tilde{m}^2 F_0^2+m^2 G_0^2\right)R^2}{F_0^2+G_0^2} \Bigg]\,,\nonumber\\
\omega_B^2 &=& \Bigg[ k^2 +\left(\frac{3\,\left(G_0 F_0' -F_0 G_0'\right)^2}{2\,\left(F_0^2+G_0^2\right)^2} + \frac{3\,F_0^2 G_0^2 \left(\Sigma_0' -\tilde{\Sigma}_0'\right)^2}{2\,\left(F_0^2+G_0^2\right)^2 }\right)\left(1-\sqrt{1+\Delta M^2}\right)+\frac{\left(\tilde{m}^2 F_0^2+m^2 G_0^2\right)R^2}{F_0^2+G_0^2} \Bigg]\,,
\end{eqnarray}
and for the pairs of Higgses and longitudinal vectors, we have, respectively,
\begin{eqnarray}
\omega_C^2 &=& \frac{g_3^2}{6} \left(F_0^2+G_0^2\right) + \left\{ \frac{g_3^2}{3} \left(F_0 F_1 +G_0 G_1 \right)\right\}  \nonumber\\
&&+ \Bigg[  k^2  + \frac{3\,\left(F_0^2 \Sigma_0'+ G_0^2 \tilde{\Sigma}_0'\right)^2}{\left(F_0^2+G_0^2\right)^2 }  +\frac{\left(m^2 F_0^2+\tilde{m}^2 G_0^2\right)R^2}{F_0^2+G_0^2}
\nonumber\\
&&\quad\quad+\frac{g_3^2}{6}\left(F_1^2+G_1^2+2\,F_0 F_2 +2\,G_0 G_2 \right)-\frac{g_3^2R^2}{6}\left(\frac{4\,\delta m_2^2}{g_2^2}+\frac{3\,\left(\delta m_3^2+\delta m_4^2\right)}{g_3^2}\right)
 \Bigg]\,,\nonumber
\end{eqnarray}
\begin{eqnarray}
\omega_D^2 &=& \frac{g_3^2}{6} \left(F_0^2+G_0^2\right) + \left\{ \frac{g_3^2}{3} \left(F_0 F_1 +G_0 G_1 \right)\right\}  \nonumber\\
&&+ \left[  k^2 +\frac{g_3^2}{6}\left(F_1^2+G_1^2+2\,F_0 F_2 +2\,G_0 G_2 \right)-\frac{g_3^2R^2}{6}\left(\frac{4\,\delta m_2^2}{g_2^2}+\frac{3\,\left(\delta m_3^2+\delta m_4^2\right)}{g_3^2}\right)\right]\,,\nonumber\\
\end{eqnarray}
where, as before, we identified the longitudinal vector by comparing the frequencies to the mass of $A_i^{(\lambda_6)}$ and $A_i^{(\lambda_7)}$ (\ref{transact}) at the given order in $\epsilon$. In the above, for later convenience, we defined the dimensionless quantity,
\begin{equation}
\Delta M \equiv \frac{\left(3\,\delta m_3^2-\delta m_4^2\right)R^2\,\left(F_0^2+G_0^2\right)^2}{3\left[\left(G_0 F_0'-F_0 G_0'\right)^2+F_0^2G_0^2 \left(\Sigma_0'-\tilde{\Sigma}_0'\right)^2\right]}\,.
\end{equation}
In the degenerate mass limit $\delta m_i =0$, $F_0=F$, $G_0=G$, $\Sigma_0=\Sigma$, $\tilde{\Sigma}_0=\tilde{\Sigma}$, the eigenfrequencies of the system reduce to two copies of the ones of the coupled system in the four field toy model of \cite{Gumrukcuoglu:2008fk}. We will see that this correspondence goes even further.

The eigenvectors of the light modes read (see Appendix \ref{AppE})
\begin{equation}
\begin{array}{ll}
V_{j,1} = \frac{1}{\sqrt{1+a_j^2+b_j^2}}\left(\begin{array}{r} 
 \left(a_j\,{\bf w}_1+b_j\,{\bf w}_2\right)\,\cos\alpha_j+{\bf w}_2 \sin\alpha_j 
\\ 
 \left(a_j\,{\bf w}_1+b_j\,{\bf w}_2\right)\,\sin\alpha_j-{\bf w}_2 \cos\alpha_j 
\end{array}\right) \,,
&\\
& \quad\quad (j= A,\,B)\\
V_{j,2} = \frac{1}{\sqrt{1+a_j^2+b_j^2}}\left(\begin{array}{r} 
- \left(a_j\,{\bf w}_1+b_j\,{\bf w}_2\right)\,\sin\alpha_j + {\bf w}_2 \cos\alpha_j 
\\ 
\left(a_j\,{\bf w}_1+b_j\,{\bf w}_2\right)\,\cos\alpha_j +{\bf w}_2 \sin\alpha_j
\end{array}\right) \,,
&\\
\end{array} 
\label{eveclight}
\end{equation}
where the coefficients are
\begin{eqnarray}
a_A &\equiv& \frac{\nu_-^2}{2\,\left(\frac{F_0'}{F_0} - \frac{G_0'}{G_0}\right) \left(\Sigma_0'-\tilde{\Sigma}_0'\right)}
\,,\quad
a_B\equiv -\frac{\nu_+^2}{2\,\left(\frac{F_0'}{F_0} - \frac{G_0'}{G_0}\right) \left(\Sigma_0'-\tilde{\Sigma}_0'\right)}\,,
\nonumber\\
b_A &=&b_B \equiv -\frac{\left(\frac{F_0'}{F_0} - \frac{G_0'}{G_0}\right)^2 + \left(\Sigma_0'-\tilde{\Sigma}_0'\right)^2}{2\,\left(\frac{F_0'}{F_0} - \frac{G_0'}{G_0}\right) \left(\Sigma_0'-\tilde{\Sigma}_0'\right)}\Delta M\,,
\end{eqnarray}
with definitions,
\begin{equation}
\nu_{\pm} \equiv \sqrt{\mu^2 \pm \left[\left(\frac{F_0'}{F_0} - \frac{G_0'}{G_0}\right)^2 - \left(\Sigma_0'-\tilde{\Sigma}_0'\right)^2\right]}\,,\quad\quad
\mu \equiv \sqrt{\left(\frac{F_0'}{F_0} - \frac{G_0'}{G_0}\right)^2 + \left(\Sigma_0'-\tilde{\Sigma}_0'\right)^2}\,\left(1+\Delta M^2\right)^{1/4}\,.
\end{equation}
The four dimensional vectors ${\bf w}_1$ and ${\bf w}_2$ in (\ref{eveclight})  are normalized and orthogonal eigenvectors of matrix $A$ defined in (\ref{omform6}). These can be written as combinations of (\ref{evecA}) through
\begin{equation}
{\bf w}_1 = \frac{1}{\sqrt{2}}\,({\bf w}_A+{\bf w}_B) \,,\quad {\bf w}_2 = \frac{1}{\sqrt{2}}\,(-{\bf w}_A+{\bf w}_B) \,.
\end{equation}
As for the heavy modes, the eigenvectors $V_{C,1}$, $V_{C,2}$, $V_{D,1}$ and $V_{D,2}$ in (\ref{evecsyd}) are also valid for this system.

The calculation of the matrix $\Gamma$ is then straightforward, although the expressions are far from simple. Formally, its structure is as follows
\begin{equation}
\Gamma = \left(
\begin{array}{rrrrrrrr}
0 & \Gamma_{12} & \Gamma_{13} & \Gamma_{14} & \Gamma_{15} &\Gamma_{16} & 0 &0\\
  & 0 & -\Gamma_{14} & \Gamma_{13} & -\Gamma_{16} & \Gamma_{15} & 0 & 0\\
&& 0 & \Gamma_{34} & \Gamma_{35} & \Gamma_{36} & 0 & 0\\
&&& 0 & -\Gamma_{36}  & \Gamma_{35} & 0& 0\\
&&&&0 & \Gamma_{56} &0 &0\\
&&&&&0 &0 &0\\
&&&&&& 0 & \Gamma_{78}\\
&&&&&&& 0 
\end{array}
\right)\,\epsilon+\mathcal{O}(\epsilon^2)\,,
\end{equation}
with a total of ten independent components. We find that, as in the previous system, it is not possible to remove the non-adiabatic mixing by rotating the eigenvectors of degenerate states. In general, from the suppression arguments made in Section \ref{sec:formalism},  we expect production due to mixings between the light modes and the Higgses. As every physical mode has an identical copy, one way to simplify the system is by decoupling the copies into two independent systems.

We first remove the mixings between the copies of the heavy modes, which are due to the components $\Gamma_{56}$ and $\Gamma_{78}$. These vanish by choosing the rotation parameters
\begin{equation}
\alpha_C =\alpha_{C 0} \,,\quad \alpha_D =\alpha_{D 0}\,,
\label{solCD}
\end{equation}
with constant $\alpha_{C 0 }$ and $\alpha_{D0}$. We now simplify the mixings between different modes. For any two sets of distinct eigenvalues, there corresponds four components of $\Gamma$ matrix. Our goal is to keep only two of these, and remove the other two. For the mixing between modes $(1,2)$ and $(5,6)$, we choose to solve $\Gamma_{15}=0$  which is an algebraic equation, with the solution
\begin{equation}
 \tan ( \alpha_A-\alpha_C) = \left(\frac{\frac{F_0'}{F_0}-\frac{G_0'}{G_0}}{\Sigma_0'-\tilde{\Sigma}_0'}\right)\,\left(\frac{\Delta M}{\sqrt{1+\Delta M^2}+1}\right)\,.
\end{equation}
In a similar way, we remove a pair of components corresponding to the mixings between the modes $(3,4)$ and $(5,6)$, by requiring $\Gamma_{36}=0$, solved by
\begin{equation}
 \tan ( \alpha_B-\alpha_C) = \left(\frac{\Sigma_0'-\tilde{\Sigma}_0'}{\frac{F_0'}{F_0}-\frac{G_0'}{G_0}}\right)\,\left(\frac{\Delta M}{\sqrt{1+\Delta M^2}+1}\right)\,.
\end{equation}
Now, we have fixed all four of the rotation parameters up to integration constants $\alpha_{C0}$ and $\alpha_{D0}$. The logical progression of this procedure is to set $\Gamma_{12}=0$ and $\Gamma_{34}=0$, i.e. to remove the mixing between the two copies of the light modes. This, in principle should give two algebraic equations for $\alpha_{C0}$ and $\alpha_{D0}$. 
However, these equations require time dependent solutions, inconsistent with (\ref{solCD}). Although we had previously argued that $\Gamma_{12}$ and $\Gamma_{34}$ will not result in particle production due to adiabatic behavior of these specific mixings, they will still be responsible for converting produced quanta into their twin copy. So our goal of decoupling the two copies cannot be realized in this setting. However, relaxing the different mass requirement simplifies the problem considerably. From here on, we will assume $\Delta M \ll 1$ and use the above choices for the rotation parameters and expand the remaining six $\Gamma$ matrix components in series in $\Delta M$. At order $\mathcal{O}(\Delta M^0)$ and $\mathcal{O}(\epsilon)$, the remaining components of $\Gamma$ are
\begin{eqnarray}
\Gamma_{12} &=& \mathcal{O}(\Delta M)\,,\quad\quad\quad
\Gamma_{14} = \mathcal{O}(\Delta M)\,,\quad\quad\quad
\Gamma_{16} = \mathcal{O}(\Delta M)\,,\quad\quad\quad
\Gamma_{34} = \mathcal{O}(\Delta M)\,,\nonumber\\
\Gamma_{13} &=& {\rm sgn}\left[\left(\frac{F_0'}{F_0}-\frac{G_0'}{G_0}\right)\left(\Sigma_0'-\tilde{\Sigma}_0'\right)\right]
\left(\frac{\left(m^2-\tilde{m}^2\right) R^2\left(\Sigma_0'-\tilde{\Sigma}_0'\right)}{\left(\frac{F_0'}{F_0}-\frac{G_0'}{G_0}\right)^2+\left(\Sigma_0'-\tilde{\Sigma}_0'\right)^2} - \frac{F_0^2\Sigma_0'+G_0^2\tilde{\Sigma}_0'}{F_0^2+G_0^2}\right)
+\mathcal{O}(\Delta M^2)\,,\nonumber\\
\Gamma_{35} &=& {\rm sgn}\left[\Sigma_0'-\tilde{\Sigma}'_0\right]\,\frac{F_0 G_0}{F_0^2+G_0^2}\, \sqrt{\left(\frac{F_0'}{F_0}-\frac{G_0'}{G_0}\right)^2 + \left(\Sigma_0'-\tilde{\Sigma}_0'\right)^2}+\mathcal{O}(\Delta M^2)
\end{eqnarray}
In this approximation, the modes $(1,3,5)$ and $(2,4,6)$ decouple from each other and form two copies of exactly the same system, which resembles to the coupled system of four field toy model with $U(1)$ symmetry in \cite{Gumrukcuoglu:2008fk}. 
By comparing the eigenfrequencies, the $\Gamma$ matrix components of the two problems are equivalent by the exchange $(\Gamma_{13},\,\Gamma_{24} )\rightarrow \Gamma_{23}^{\rm toy ~model}$ and $(\Gamma_{35},\,\Gamma_{46}) \rightarrow -\Gamma_{13}^{\rm toy ~model}$, with $g_3^2 \rightarrow 3\,e^2/2$. \footnote{The overall signs of $\Gamma$ matrix in \cite{Gumrukcuoglu:2008fk} can be recovered by doing a rotation on $(1,2)$ by an angle $\left(1-{\rm sgn}[F_0'/F_0- G_0'/G_0]\right)\pi/2$ and on $(3,4)$ by angle $\left(1-{\rm sgn}[\Sigma_0'- \tilde{\Sigma}'_0]\right)\pi/2$.} It should also be noted that the sub-leading terms in the eigenmasses of the heavy modes do not match exactly with the ones in the toy model, but being corrections to heavy mode mass, they will have a suppressed effect on the final result. However, as mentioned above, in the limit of degeneracy, the two problems give exactly the same equations at the given order of $\epsilon$ expansion.

\section{Doubly degenerate system and uniqueness of occupation numbers}\label{AppE}

Here, we outline the diagonalization of the matrix form encountered in the last two subsections of Appendix \ref{AppD}. The eigenproblem for the $2\,N \times 2\,N$ matrix of form
\begin{equation}
\Omega_4^2+K_4^TK_4 = \left( 
\begin{array}{ll}
A & B\\
-B & A
\end{array}\right)\,,
\label{matrix2n}
\end{equation}
where $A$ and $B$ are $N\times N$ real matrices with $A^T = A$ and $B^T = -B$, is equivalent to the eigenproblem of $N\times N$ Hermitian matrix, 
\begin{equation}
\left(A -i\,B\right)\cdot \left(V_R+i\,V_I\right) = \lambda \left(V_R + i \,V_I\right)\,,
\end{equation}
where lambda is an eigenvalue, $V_R$ and $V_I$ are the real and imaginary parts of the corresponding eigenvector. The above equation can be written in the form,
\begin{equation}
\left( 
\begin{array}{ll}
A & B\\
-B & A
\end{array}\right) \left(\begin{array}{l} V_R \\ V_I \end{array}\right) = \lambda \,\left(\begin{array}{l} V_R \\ V_I \end{array}\right)
\,.
\end{equation}
The phase freedom in the Hermitian problem, translates to the real case as an $SO(2)$ symmetry on the eigenvectors. As there is no way to distinguish between the pairs of the eigenvalues, the eigenvectors can be rotated into one another.

The problem in the main text is further simplified in the $\epsilon$ expansion. In both cases where we have the form (\ref{matrix2n}), we have $B = \mathcal{O}(\epsilon^2)$, and as a consequence, at zero order in $\epsilon$, both vectors $V_R$ and $V_I$ are also eigenvectors of the matrix $A$ with same eigenvalue. However, the matrix $A$ has additional degeneracies at zero order, so the eigenvectors have another rotational degree of freedom, which can be removed by solving the second order eigenvalue equations. In general, the eigenvectors of $\Omega_4^2+K_4^TK_4$ can be written as
\begin{equation}
V_1 = \frac{1}{\sqrt{1+a^2+b^2}}\left(\begin{array}{r} 
 \left(a\,{\bf w}_1+b\,{\bf w}_2\right)\,\cos\alpha+{\bf w}_2 \sin\alpha 
\\ 
 \left(a\,{\bf w}_1+b\,{\bf w}_2\right)\,\sin\alpha-{\bf w}_2 \cos\alpha 
\end{array}\right) \,,\quad
V_2 = \frac{1}{\sqrt{1+a^2+b^2}}\left(\begin{array}{r} 
- \left(a\,{\bf w}_1+b\,{\bf w}_2\right)\,\sin\alpha + {\bf w}_2 \cos\alpha 
\\ 
\left(a\,{\bf w}_1+b\,{\bf w}_2\right)\,\cos\alpha +{\bf w}_2 \sin\alpha
\end{array}\right) \,,
\label{vecform}
\end{equation}
where ${\bf w}_1$, ${\bf w}_2$ are orthonormal eigenvectors of $A$ corresponding to the same eigenvalue, $\alpha$ is the rotation parameter arising from the double degeneracy of the $\Omega_4^2+K_4^TK_4$ matrix. The coefficients $a$, $b$ can be determined uniquely through the second order eigenvalue problem. In the case where there is no degeneracy in $A$ at zero order, the above eigenvectors can be written in a much simpler form
\begin{equation}
V_1 = \left(\begin{array}{r} {\bf w}\,\cos\alpha  \\ {\bf w}\,\sin\alpha \end{array}\right) \quad\,,\quad
V_2 = \left(\begin{array}{r} -{\bf w}\,\sin\alpha  \\ {\bf w}\,\cos\alpha \end{array}\right) \quad\quad\,,
\label{vecformndeg}
\end{equation}
where ${\bf w}$ is the unique eigenvector of $A$ corresponding to the eigenvalue.

The last point of this section is to show that the occupation numbers are independent of the choice of rotation parameters.
If matrix $\xi$ diagonalizes $\Omega_4^2+K_4^TK_4$, with  $\xi^T \,(\Omega_4^2+K_4^TK_4)\xi = {\rm diag}(\omega^2_i)$, and is arranged such that the $(2m-1)$th and $(2m)$th columns correspond to eigenvectors of the $m$th distinct eigenvalue, then $\tilde{\xi}=\xi \,\mathcal{R}$ also diagonalizes it, provided that
\begin{eqnarray}
\mathcal{R}_{ij} &=& \delta_{ij} \quad\,,(i,j \neq 2m-1, 2m)\,,\nonumber\\
\mathcal{R}_{2m-1,2m-1} &=& \mathcal{R}_{2m,2m} = \cos\theta \quad\,,
\mathcal{R}_{2m-1,2m} =-\mathcal{R}_{2m,2m-1} = \sin\theta\,,
\end{eqnarray}
which rotates the eigenvectors corresponding to the $m$th distinct eigenvalue, with $m = 1,..N$. There are $N$ such independent rotations one can apply, under which the matrix $\Gamma$ (\ref{gamfin}) becomes
\begin{equation}
\tilde{\Gamma} = \mathcal{R}^T \left(\Gamma \,\mathcal{R} +\mathcal{R}'\right)\,.
\end{equation}
Here and below, an overtilde denotes the quantity calculated using the rotated eigenvectors. Similarly, the transformation laws for the matrices $I$ and $J$ (\ref{eqij}) are
\begin{equation}
\tilde{I} = \mathcal{R}^T \left(I \,\mathcal{R} +\mathcal{R}'\right)\,,\quad
\tilde{J} = \mathcal{R}^T\, J \,\mathcal{R}\,.
\end{equation}
Finally, the Bogolyubov equations will have the same form as in (\ref{bogoleq}),
\begin{equation}
\tilde{\alpha}' = \left(-i\,\omega-\tilde{I}\right)\tilde{\alpha} + \left(\frac{\omega'}{2\,\omega}-\tilde{J}\right)\tilde{\beta}\,,\quad\quad
\tilde{\beta}' = \left(i\,\omega-\tilde{I}\right)\tilde{\beta} + \left(\frac{\omega'}{2\,\omega}-\tilde{J}\right)\tilde{\alpha}\,,
\end{equation}
where we defined $\tilde{\alpha}\equiv \mathcal{R}^T \alpha$ and $\tilde{\beta} \equiv \mathcal{R}^T \beta$. Of course, as there is arbitrariness in the definition of the eigenstates, the occupation numbers of a given rotated state will vary with the rotation angle. However, the physical quantity, which is the total occupation number for the $m$th distinct eigenvalue is invariant of the choice of the rotation parameters
\begin{equation}
\tilde{n}_m = (\tilde{\beta}^\star \tilde{\beta}^T)_{2m-1,2m-1}+(\tilde{\beta}^\star \tilde{\beta}^T)_{2m,2m} =(\mathcal{R}^T \beta^\star \beta^T \mathcal{R})_{2m-1,2m-1}+(\mathcal{R}^T \beta^\star \beta^T \mathcal{R})_{2m,2m}  = n_m\,.
\end{equation}

\end{document}